\documentclass[a4paper]{aa}

\usepackage{graphicx,natbib}
\usepackage{txfonts}
\usepackage{subfigure}
\usepackage{dblfloatfix}
\usepackage{fixltx2e}
\usepackage[usenames,dvipsnames]{xcolor}

\newcommand{\kms}{km\,s$^{-1}$}
\newcommand{\inv}{{\sc Invers13}}

\begin{document}

\title{How reliable is Zeeman Doppler Imaging without simultaneous temperature reconstruction?}

\author{L.~Ros\'{e}n
   \and O.~Kochukhov}

\institute{Department Physics and Astronomy, Uppsala University, Box 516, 751 20 Uppsala, Sweden}

\date{Received 9 July 2012 / Accepted 27 September 2012}

\titlerunning{How reliable is ZDI of cool active stars?}

\abstract
{%
}
{The goal of this study is to perform numerical tests of Zeeman Doppler Imaging (ZDI) to asses whether correct reconstruction of magnetic fields is at all possible without taking temperature into account for stars in which magnetic and temperature inhomogeneities are spatially correlated.}
{We used a modern ZDI code employing a physically realistic treatment of the polarized radiative transfer in all four Stokes parameters. We generated artificial observations of isolated magnetic spots and of magnetic features coinciding with cool temperature spots and then reconstructed magnetic and temperature distributions from these data.}
{Using Stokes $I$ and $V$ for simultaneous magnetic and temperature mapping for the star with a homogeneous temperature distribution yields magnetic field strengths underestimated by typically 10--15\% relative to their true values. When temperature is kept constant and Stokes $I$ is not used for magnetic mapping, the underestimation is 30--60\%. At the same time, the strength of magnetic field inside cool spots is underestimated by as much as 80--95\% and the spot geometry is also poorly reconstructed when temperature variations are ignored. On the other hand, the inversion quality is greatly improved when temperature variations are accounted for in magnetic mapping. The field strength is underestimated by 40--70\% for the radial and azimuthal spots and by 70--80\% for the meridional spots. Inversions still suffer from significant crosstalk between radial and meridional fields at low latitudes. The azimuthal field component proves to be most robust since it suffers the least from crosstalk. When using all four Stokes parameters crosstalk is removed. In that case, the reconstructed field strength inside cool spots is underestimated by 30--40\% but the spot geometry can be recovered very accurately compared to the experiments with circular polarization alone.}
{Reliable magnetic field reconstruction for a star with high-contrast temperature spots is essentially impossible if temperature inhomogeneities are ignored. A physically realistic line profile modeling method, which simultaneously accounts for both types of inhomogeneities, is required for meaningful ZDI of cool active stars.}


\maketitle

\section{Introduction}
\label{intro}

Stellar magnetic fields vary widely in strength and configuration. There are many types of magnetic field generation mechanisms, but not even the magnetic field generation in the Sun is completely understood. To be able to understand the different types of magnetic fields and their origin we first need to know how the fields are structured on the stellar surfaces and how their configurations evolve with time. 

Doppler Imaging (DI) has long been used to reconstruct 2D maps of the distribution of temperature or chemical abundance on stellar surfaces \citep{Vogt1987,Piskunov1990,Kochukhov2004}. This technique uses rotationally modulated unpolarized (Stokes $I$) line profiles at different rotational phases to reconstruct a map of inhomogeneities on the stellar surface. Temperature spots on cool active stars will show up in the line profiles as intensity variations, or bumps. The position of the temperature spot on the star will determine the position of the bump within the line profile. Reversely, the position of the bump can be used to infer the longitude of the temperature spot \citep{Vogt1983}. By analyzing line profiles at different rotational phases, the latitude of the temperature spot can also be determined. If the spot lies close to the rotational equator, the bump will show up in the far blue wing of the profile and then move across the line. On the other hand, a temperature spot near a visible pole will only produce a distortion close to the middle of the line profile. 

If high-resolution \textit{circularly} polarized (Stokes $V$) observations are included in DI, a tool for magnetic field reconstruction, called Zeeman Doppler Imaging (ZDI) \citep{Semel1989}, is obtained. It uses the same basic principles as conventional DI, but tries to solve a far more challenging problem of interpreting polarization signatures inside spectral lines in terms of the surface distribution of vector magnetic fields.

Nowadays ZDI is a widely used technique \citep{Donati2003,Petit2004,Catala2007}, applied to many different types of hot and cool magnetic stars. This widespread usage requires testing validity of the method. Several studies demonstrated that ZDI is useful, although it has its limitations. For example, \citet{Brown91} found that ZDI using only Stokes $V$ could reproduce a spotted magnetic field structure more accurately than a large dipole-like structure. They did however chose to dissociate temperature spots and magnetic spots by reconstructing a brightness map for a star with temperature spots only, and, conversely, by only reconstructing the magnetic field for a star with magnetic spots but without any temperature inhomogeneities. They also used an analytical line formation model assuming a plane parallel Milne-Eddington model atmosphere and applied inversions to simulated observations of a single magnetically sensitive spectral line. Each magnetic map was reconstructed under the assumption of a radial field.  

Further tests carried out by \citet{Donati1997} showed that ZDI can, to some extent, reproduce the field \textit{orientation} in a spot. The method proved to be particularly effective at distinguishing between a radial or meridional field from an azimuthal field. They also found crosstalk  from radial to meridional field components and vice versa, especially at low latitudes. \citet{Donati1997} used an even simpler weak-field Gaussian analytical treatment of the line profiles, which does not allow realistic computation of the polarization profiles for magnetic stars with temperature inhomogeneities. Both \citet{Brown91} and \citet{Donati1997} constrained their inversions by the Maximum Entropy method.

Due to weakness of surface magnetic fields on cool stars, they are usually observed in Stokes $I$ and $V$ parameters only. Using an independent ZDI code employing Tikhonov regularization, \citet{Kochukhov02} investigated the difference in using all four Stokes parameters for magnetic inversions compared to using only circular polarization. In agreement with \citet{Donati1997}, they found a strong crosstalk between radial and meridional field components at low latitudes, which can be alleviated if inversions are based on the full Stokes vector observations. However, the tests by \citet{Kochukhov02} targeted early-type Ap stars with chemical spots and did not incorporate temperature inhomogeneities in magnetic field inversions nor did they try to simultaneously reconstruct the magnetic field and temperature maps. 

Summarizing, the conventional method of mapping magnetic fields in cool stars is to use Stokes $V$ line profiles only, ignoring local temperature variations which can be derived from simultaneous analysis of Stokes $I$. Previous ZDI numerical experiments also mainly focused on purely magnetic or purely temperature spots. Yet, from general theoretical considerations and observations of the Sun we expect a strong correlation of the temperature and magnetic features. It is not clear just by how much this correlation can undermine magnetic maps reconstructed from Stokes $V$ alone.

The goal of our paper is to perform a series of numerical tests to verify validity of ZDI maps obtained for stars with temperature spots and to investigate if the quality of magnetic mapping would increase if temperature inhomogeneities are taken into account in a self-consistent and physically realistic manner. The inversion code we have used is suitable for this task since it computes the Stokes line profiles self-consistently, taking into account both the local Zeeman effect \textit{and} the local temperature.       

Our paper is structured as follows. We describe our ZDI code in Sect.~\ref{zdi-code} and discuss the setup for numerical experiments in Sect.~\ref{exp}. We then present inversions from the simulated data in Sect.~\ref{res}, evaluate these results in Sect.~\ref{dis}, and summarize main conclusions of our investigation in Sect.~\ref{con}.

\section{Zeeman Doppler Imaging code}
\label{zdi-code}

In this paper we carried out ZDI numerical experiments using our new inversion code \inv. This magnetic imaging software has been developed from the {\sc Invers10} code \citep{Piskunov02,Kochukhov02}, previously applied for the reconstruction of magnetic field topologies and distributions of chemical spots on Ap stars \citep[e.g.,][]{Kochukhov10,Luftinger10}. \inv\ incorporates the full numerical treatment of the polarized radiative transfer in a realistic stellar model atmosphere. This allows us to calculate the spectra in all four Stokes parameters simultaneously and self-consistently, without relying on any of the common simplifying assumptions, such as Gaussian profiles in the weak-field limit or the Milne-Eddington atmosphere. Of particular importance for the magnetic mapping of cool stars is a self-consistent treatment of the temperature and magnetic stellar surface structures: the temperature variations are accounted for in computing the polarization profiles and, at the same time, the Zeeman splitting of spectral lines is incorporated in theoretical Stokes $I$ spectra. The approach implemented in \inv\ is superior to most other applications of ZDI, which systematically neglect the effects of temperature spots in the magnetic inversions for cool active stars. 

Calculation of the local Stokes parameter profiles by \inv\ is based on the spectral line lists extracted from the {\sc vald} database \citep{Kupka1999}. The model atmospheres for a given metallicity, surface gravity and a range of effective temperatures are adopted from the {\sc marcs} grid \citep{Gustafsson2008}. \inv\ calculates the Stokes $IQUV$ profiles and continuum intensities for an arbitrary effective temperature of a given stellar surface element by interpolating between the sets of model spectra corresponding to the three nearest points in the model atmosphere grid. This allows an accurate semi-analytical computation of the derivatives with respect to temperature. The derivatives with respect to the three magnetic field vector components are evaluated numerically, using a simple one-sided difference scheme. 

The local spectra are convolved with a Gaussian profile to take into account instrumental and radial-tangential macroturbulent broadening. Then, the Stokes spectra are Doppler-shifted, interpolated on the wavelength grid of the observed spectra and summed for each rotational phase. The resulting disk-integrated Stokes parameter profiles are normalized by the phase-dependent continuum flux and compared with observations. In calculating the goodness of fit it is essential to balance the contributions of each Stokes parameter by scaling the respective chi-square terms by the inverse of the mean amplitude of the corresponding Stokes parameter. This usually implies a 10 to 30 times higher weight for the Stokes $V$ profiles compared to Stokes $I$.

\begin{table*}[!ht]
\caption{Configuration of the magnetic and temperature spots adopted for ZDI numerical experiments.}
\label{tab1}
\centering
\begin{tabular}{ccccccc}
\hline\hline
Latitude & Longitude & Spot radius (R$_\star$)& $B_{\rm r}$ (kG) & $B_{\rm m}$ (kG) & $B_{\rm a}$ (kG) & $T_{\rm s}$ (kK)   \\
\hline
0   & 0     & 0.40 & 1.0 & 0 & 0 & 4.75  \\
20 & 90   & 0.40 & 1.0 & 0 & 0 & 4.75  \\
40 & 180 & 0.40 & 1.0 & 0 & 0 & 4.75  \\
60 & 270 & 0.40 & 1.0 & 0 & 0 & 4.75  \\
0   & 0     & 0.25 & 2.0 & 0 & 0 & 3.75  \\
20 & 90   & 0.25 & 2.0 & 0 & 0 & 3.75  \\
40 & 180 & 0.25 & 2.0 & 0 & 0 & 3.75  \\
60 & 270 & 0.25 & 2.0 & 0 & 0 & 3.75  \\
\hline 
\end{tabular}
\tablefoot{A similar setup was used for the experiments with meridional and azimuthal fields inside cool spots.}
\end{table*}

The ZDI with only Stokes $I$ and $V$ spectra available for cool active stars is an intrinsically ill-posed problem, requiring the use of regularization to reach a stable and unique solution \citep{Brown91,Piskunov02}. Reconstruction of temperature in \inv\ is stabilized using the Tikhonov method, similar to many previous DI studies of cool active stars \citep[e.g.,][]{Piskunov93}. Tikhonov regularization ensures that the code converges on a surface distribution with a minimum contrast between adjacent surface elements. For the magnetic field reconstruction, \inv\ provides a choice between either directly mapping the three magnetic vector components and applying the Tikhonov regularization individually to radial, meridional and azimuthal magnetic maps as described by \citet{Piskunov02} or expanding the field into a spherical harmonic series following the approach by \citet{Donati06}. We used the former regularization method in the ZDI numerical experiments presented in this paper because our goal was to investigate reliability of reconstruction of isolated small-scale magnetic features.

The iterative adjustment of the surface magnetic and temperature maps to match available observations and satisfy regularization constraints is accomplished by means of the modified Levenberg-Marquardt optimization \citep{Piskunov02}. This algorithm enables convergence in typically 10--20 iterations starting from a homogeneous temperature distribution and zero magnetic field. \inv\ code is optimized for execution on parallel computers using MPI libraries. For this particular study, we were running the code on a 16-CPU HP V-class 2500 server. Further details on the numerical and computational methods employed by \inv\ can be found in \citet{Kochukhov12}.

\section{Setup for numerical experiments}
\label{exp}

For the calculations in this paper we used a grid consisting of 1176 surface elements with roughly equal areas. Each element was assigned a specific value of magnetic field strength and temperature. The inclination angle $i$ of the star was set to 50$^\circ$, the projected rotational velocity was $v_{\rm e}\sin i$ = 40~\kms\ and the photospheric temperature was 5750~K. We created four circular magnetic spots at different latitudes and longitudes. By doing so we could investigate the quality of mapping as a function of spot position, similar to the experiments in previous studies \citep{Donati1997,Kochukhov02}. All spots were identical in shape and field strength, with a difference of 1~kG between the umbra and penumbra. The spots were circular, with a radius of $0.25 R_{\star}$ for the umbra and $0.4 R_{\star}$ for the penumbra. All magnetic spots also had the same field orientation: either radial, meridional or azimuthal. The surface outside the spots was assumed to be non-magnetic. 

We used two different temperature distributions. The first was homogeneous, i.e. the entire surface including the area covered by magnetic spots had a temperature equal to the photospheric temperature. This configuration of homogeneous temperature and magnetic spots, used in the first set of experiments, represents a reference for the subsequent reconstructions of fields inside cool spots. It may also correspond to a real situation for active stars with a hypothetical non-solar dynamo producing temperature structures not associated with particularly strong magnetic fields, or for low-activity stars for which temperature inhomogeneities are unresolved and only large-scale magnetic fields are detected \citep[e.g.][]{Petit2008}.

The second temperature distribution consisted of four cool temperature spots with the exact same shapes and positions as the magnetic spots. The umbra and penumbra was, respectively, 2000~K and 1000~K cooler than the surrounding photosphere. This configuration of temperature and magnetic spots was used in the second and third sets of experiments. A complete description of the surface structure setup for the calculation using radial magnetic spots can be found in Table~\ref{tab1}. 

For all the experiments we used a grid of 22 {\sc marcs} model atmospheres \citep{Gustafsson2008} with the surface gravity $\log g$\,=\,4.5 and microturbulence $\xi_t$\,=\,2~\kms. Since the temperature of the umbra was set to 3750~K and the photospheric temperature was 5750~K, the {\sc marcs} model atmosphere grid had a range in $T_{\rm eff}$ between 3000~K and 6750~K in order to cover these temperatures. We used three magnetically sensitive \ion{Fe}{i} lines at 5497.5, 5501.5 and 5506.8~\AA. To make our calculations realistic, we also included seven metal lines blending the \ion{Fe}{i} lines of interest. We extracted atomic line data from the {\sc vald} data base \citep{Kupka1999} to get information necessary for spectrum synthesis and to obtain Land\'e factors and quantum numbers required for Zeeman splitting calculations. These data were then used as input for forward calculations in order to obtain the rotationally modulated line profiles of either Stokes $I$ and $V$, or all four Stokes parameters, at 20 equally spaced rotational phases. Profiles were convolved with a Gaussian function to simulate instrumental resolution of $\lambda/\Delta\lambda = 65000$, but no noise was added to the profiles since we wanted to test the performance of ZDI under ideal conditions. That is also why we chose to use such a good phase coverage and optimal values for $v_{\rm e}\sin i$ and $i$. Other ZDI studies have investigated the effect of different phase coverage, different $v_{\rm e}\sin i$, different inclination, and different noise level \citep{Donati1997,Kochukhov02}.

The resulting line profiles were then used as an input for inverse calculations to recover the surface maps. In experiment 1 and 2 we first used only Stokes $V$ to reconstruct the magnetic field maps for each of the three magnetic field vector components. Then we used both Stokes $I$ and $V$ to reconstruct the magnetic field and temperature map. In the last experiment we used Stokes $I$, $Q$, $U$ and $V$ parameters for magnetic and temperature mapping. Below we present results of the inversions for these three sets of tests.

\begin{figure*}[!th]
\centering
\subfigure{\includegraphics[scale=0.55,angle=90]{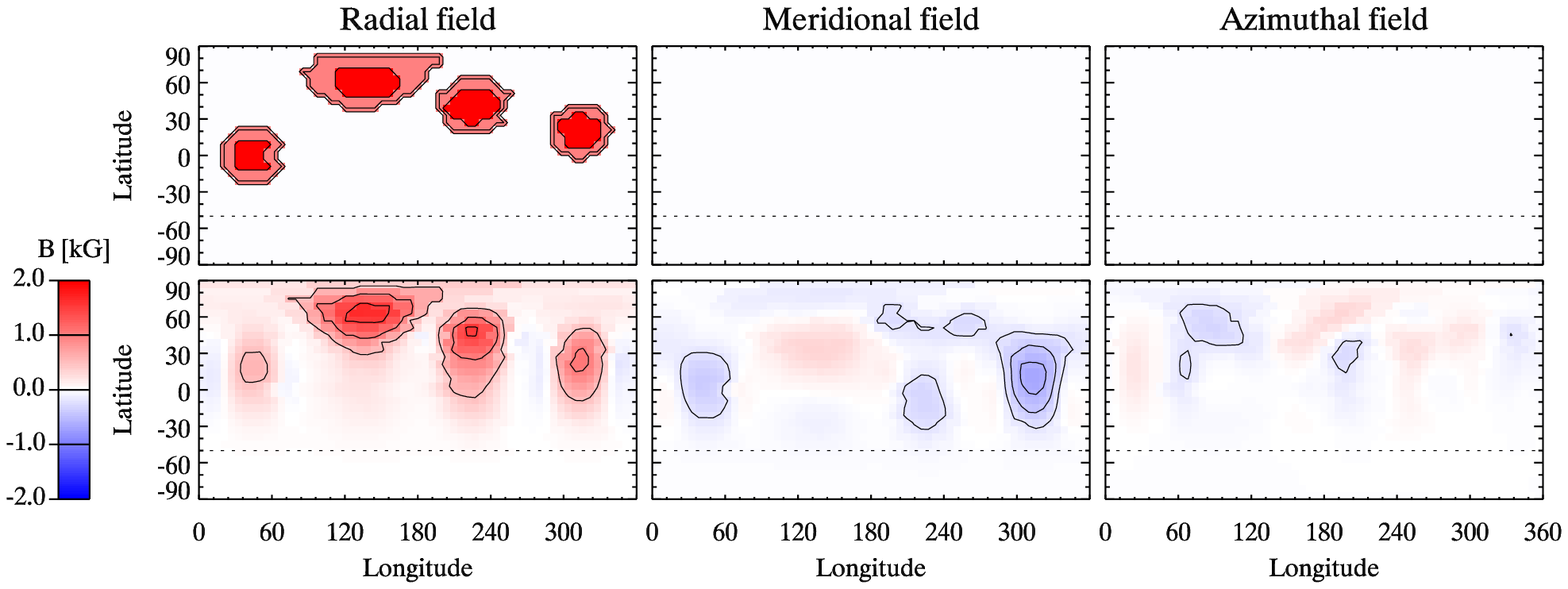}
\label{fig:subfig7}}
\centering
\subfigure{\includegraphics[scale=0.55,angle=90]{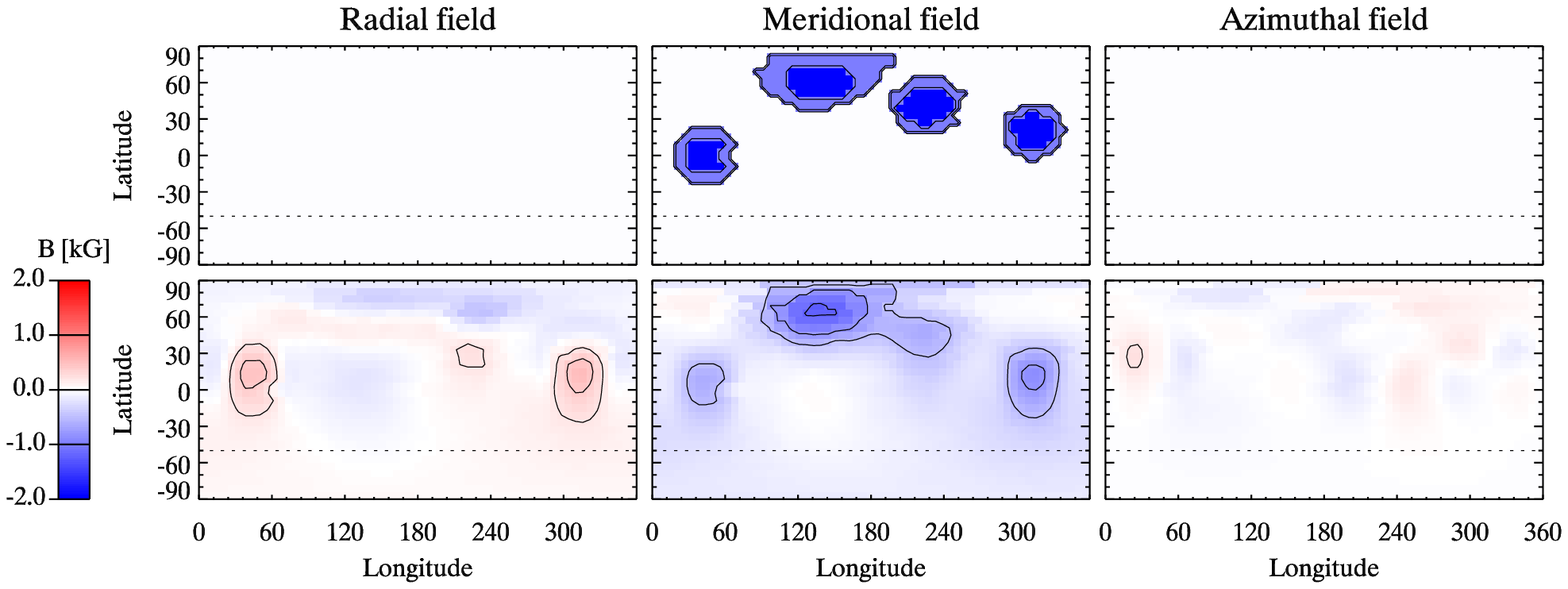}
\label{fig:subfig10}}
\centering
\subfigure{\includegraphics[scale=0.55,angle=90]{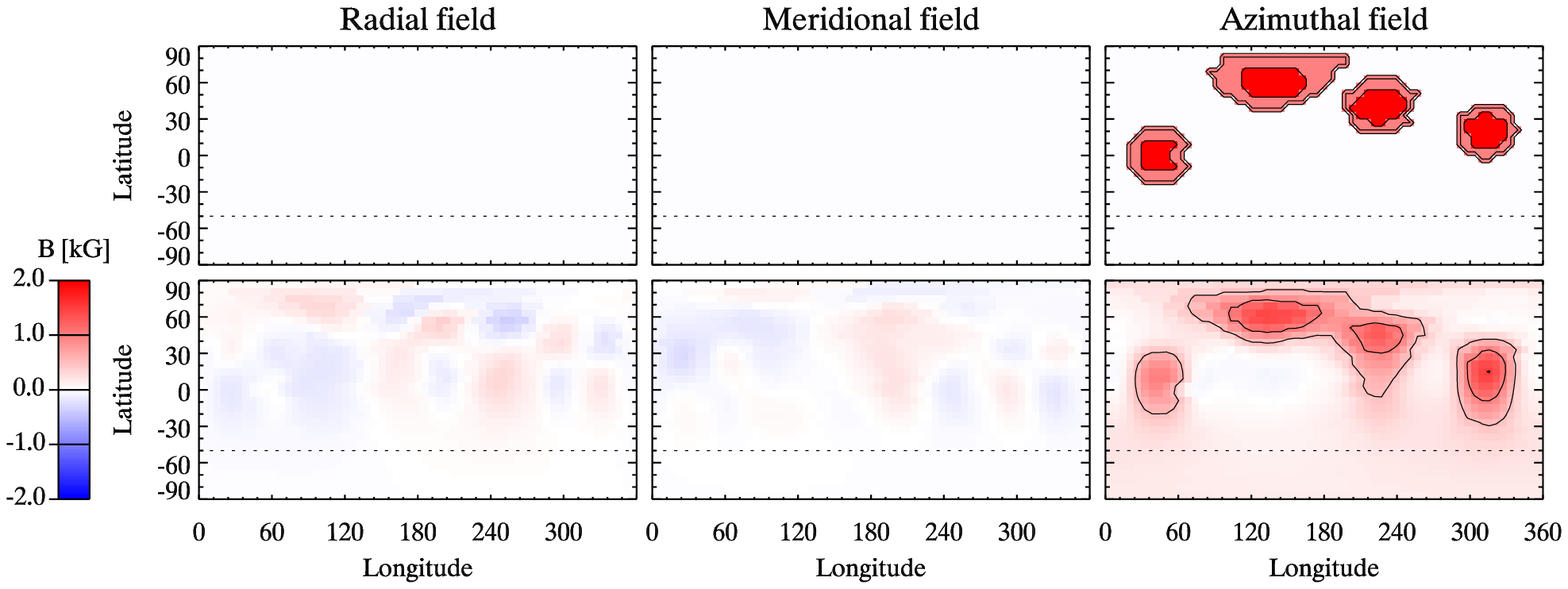}
\label{fig:subfig11}}
\caption[Optional caption for list of figures]{Initial and reconstructed magnetic field maps for the inversions with a fixed homogeneous temperature. The upper panel in each pair of rows corresponds to the true maps while the lower panel shows the reconstructed maps. These inversions were carried out using only Stokes $V$ and assuming a constant temperature.
}
\label{fig:maps1}
\end{figure*}

\begin{figure*}[!th]
\centering
\subfigure{\includegraphics[scale=0.55,angle=90]{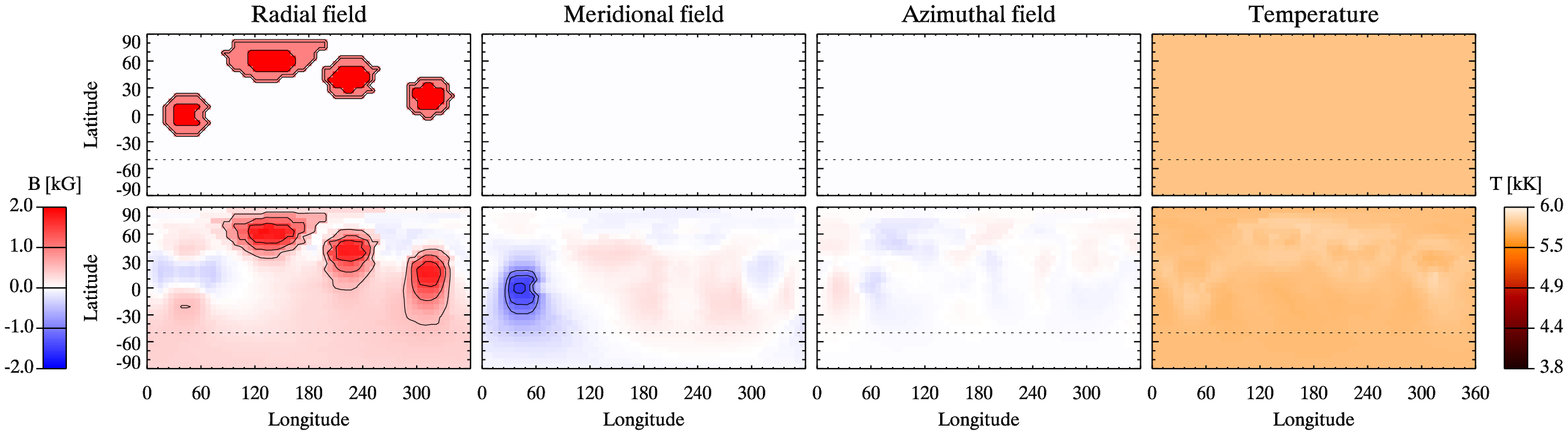}
\label{fig:subfig8}}
\subfigure{\includegraphics[scale=0.55,angle=90]{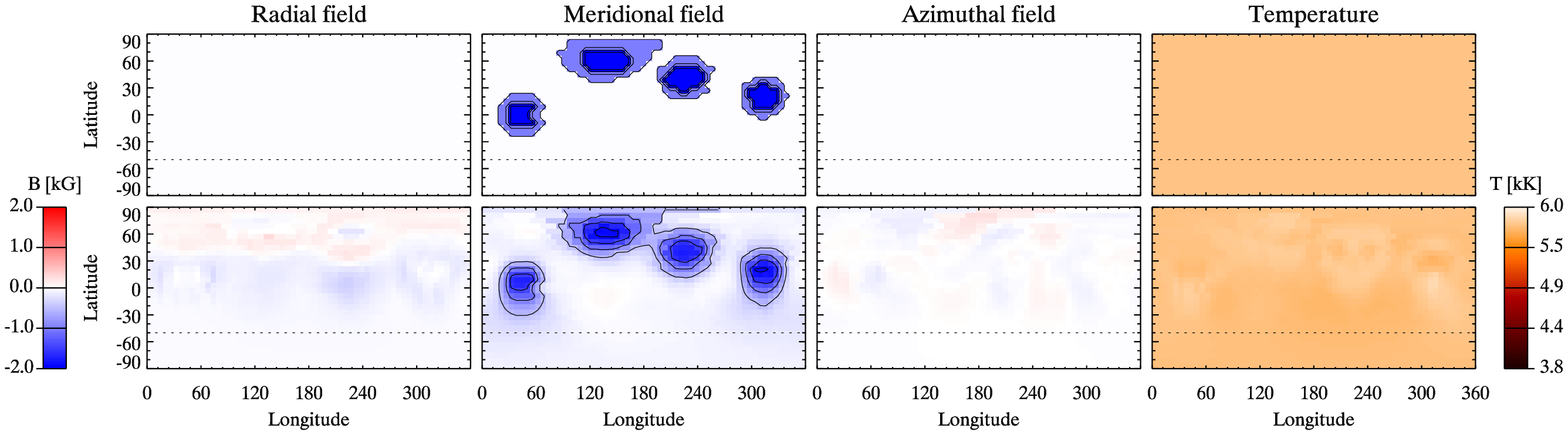}
\label{fig:subfig9}}
\subfigure{\includegraphics[scale=0.55,angle=90]{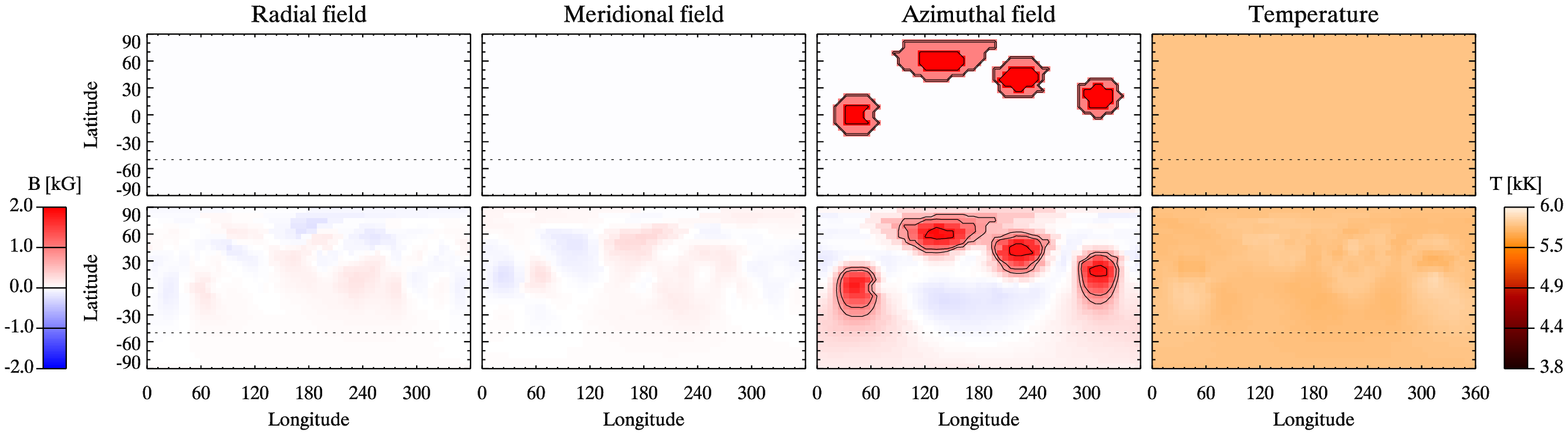}
\label{fig:subfig12}}
\caption[Optional caption for list of figures]{Same as for Fig.~\ref{fig:maps1} but here both Stokes $I$ and $V$ were used and possible temperature inhomogeneities were taken into account in self-consistent ZDI inversions.}
\label{fig:maps2}
\end{figure*} 

\begin{figure*}[!th]
\centering
\subfigure{\includegraphics[scale=0.55,angle=90]{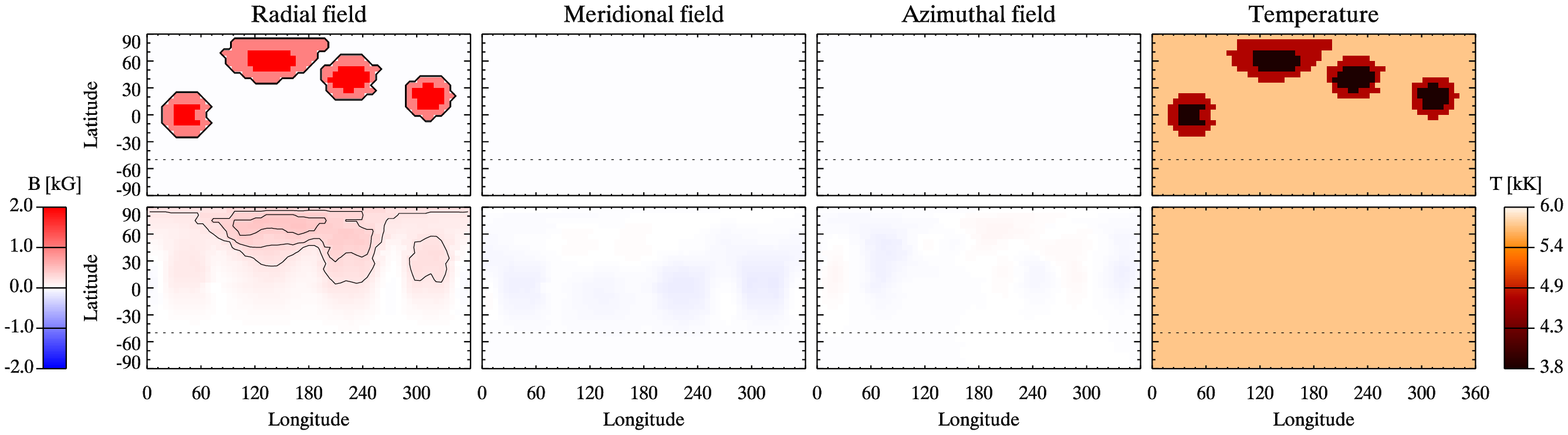}
\label{fig:subfig1}}
\centering
\subfigure{\includegraphics[scale=0.55,angle=90]{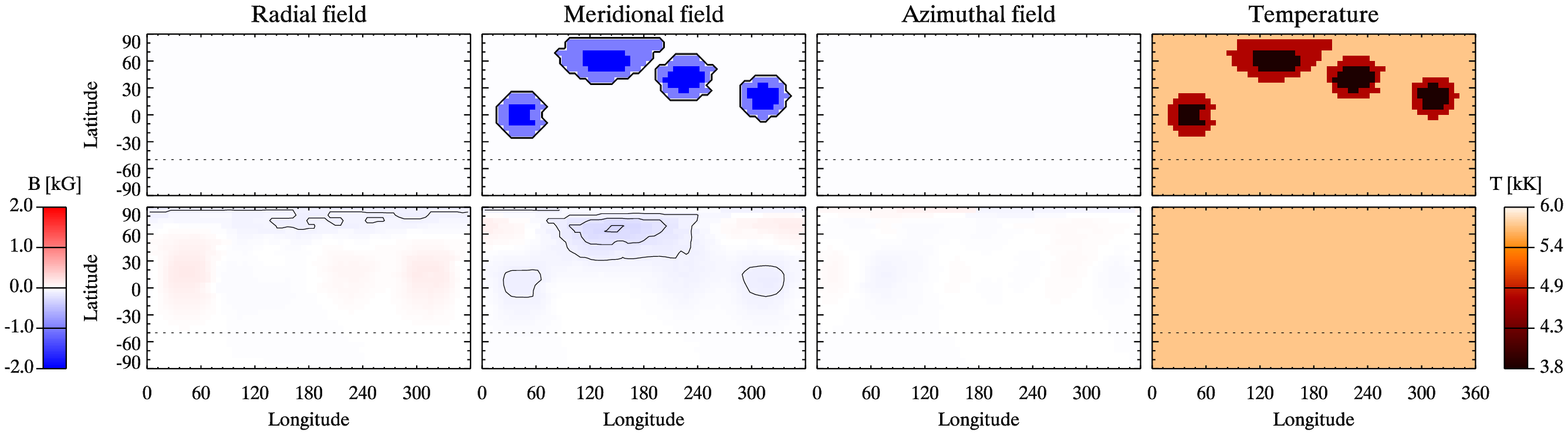}
\label{fig:subfig3}}
\centering
\subfigure{\includegraphics[scale=0.55,angle=90]{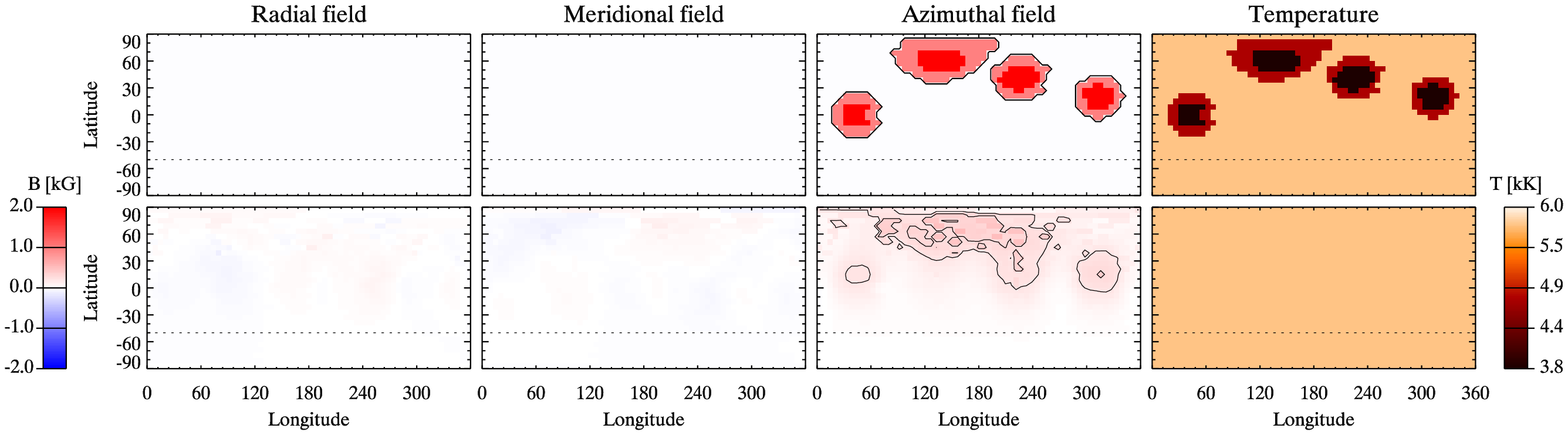}
\label{fig:subfig5}}
\caption[Optional caption for list of figures]{Same as for Fig.~\ref{fig:maps1} but for the inversions with the magnetic spots coinciding with low-temperature spots.
These inversions were carried out using only Stokes $V$ and assuming a constant temperature.
}
\label{fig:maps3}
\end{figure*}

\begin{figure*}[!th]
\centering
\subfigure{\includegraphics[scale=0.55,angle=90]{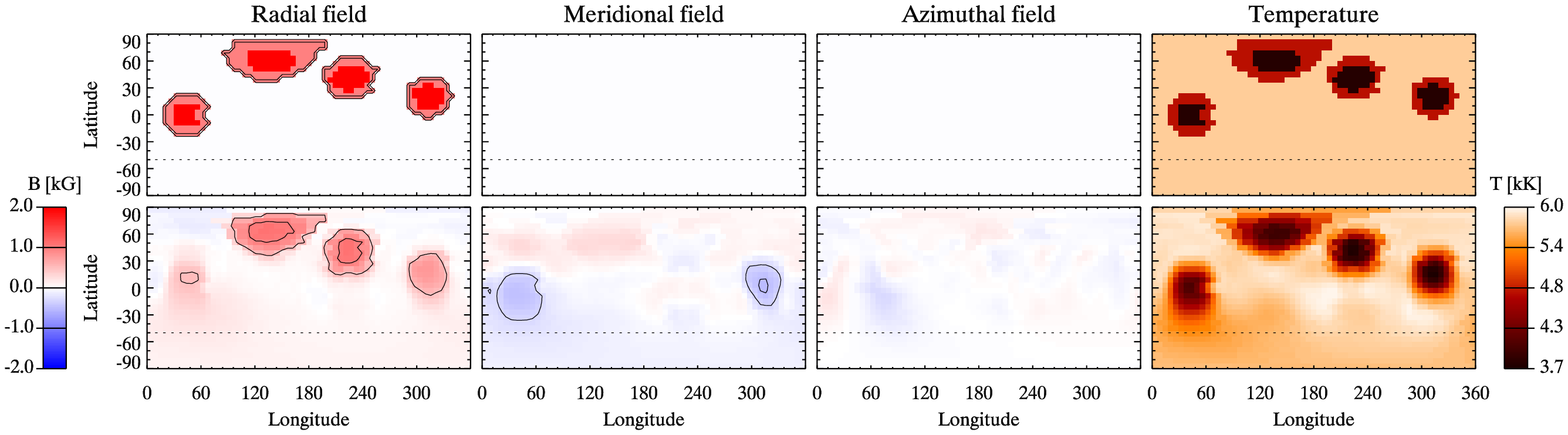}
\label{fig:subfig2}}
\centering
\subfigure{\includegraphics[scale=0.55,angle=90]{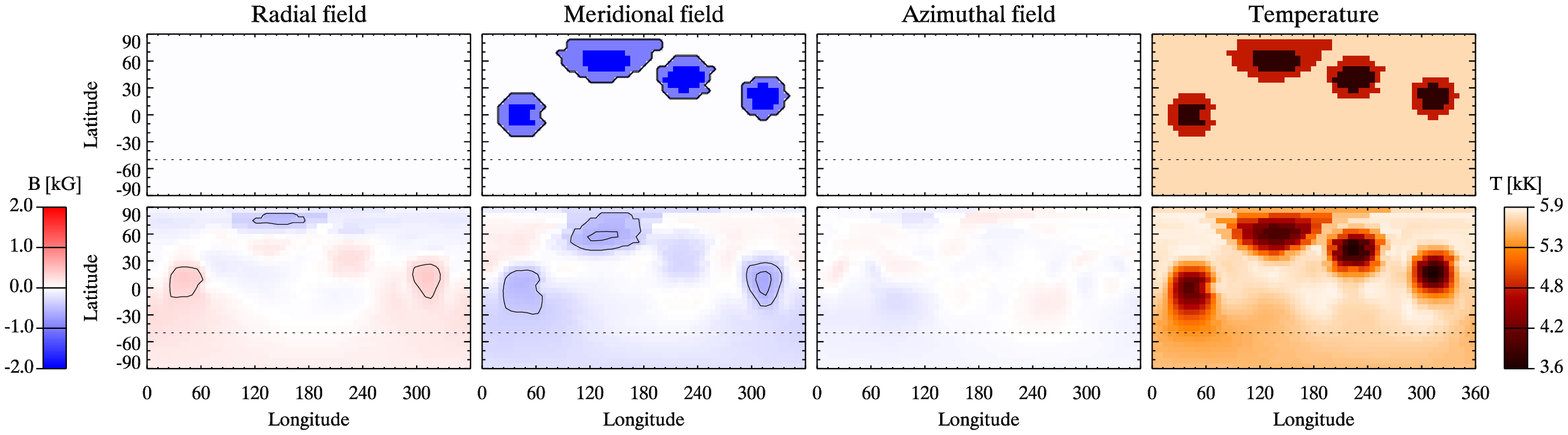}
\label{fig:subfig4}}
\centering
\subfigure{\includegraphics[scale=0.55,angle=90]{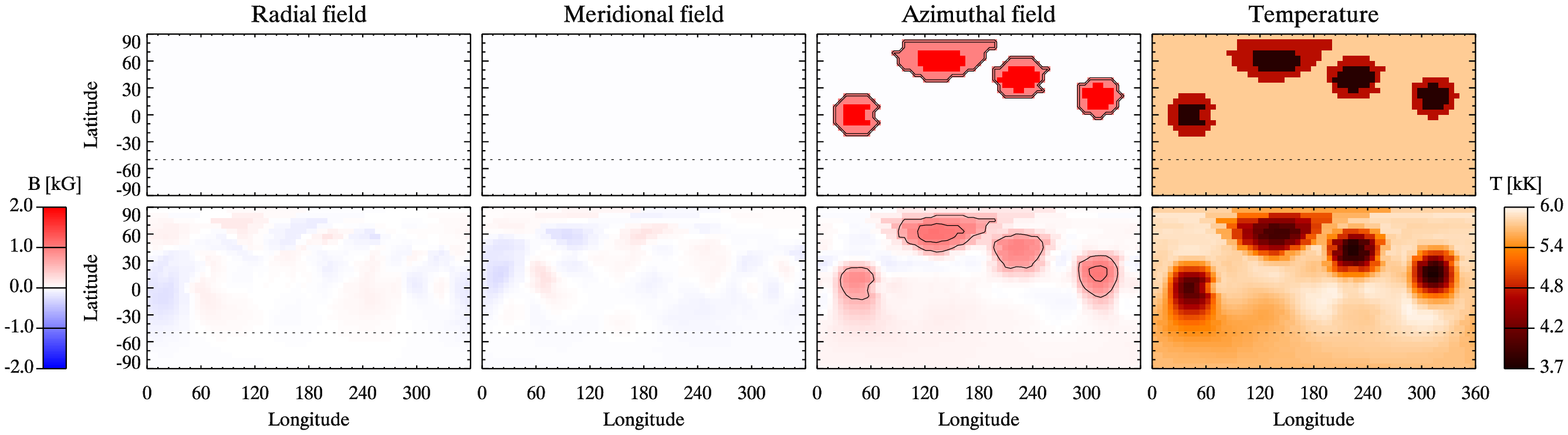}
\label{fig:subfig6}}
\caption[Optional caption for list of figures]{Same as for Fig.~\ref{fig:maps3} but here both Stokes $I$ and $V$ were used and  temperature inhomogeneities were taken into account in self-consistent ZDI inversions.}
\label{fig:maps4}
\end{figure*}

\section{Results}
\label{res}

In the following sections we discuss results of the simulated ZDI inversions for purely magnetic spots using Stokes $IV$, for magnetic field concentrations coinciding with low-temperature spots by first using Stokes $IV$, and, finally, by using all four Stokes parameters for the inversions. In each case we reconstructed magnetic and temperature maps for different magnetic field vector orientations. We assess these inversion results by the visual comparison of the true and reconstructed surface distributions represented in rectangular projection maps. We also evaluate the inversions numerically, by computing by how much the magnetic field strength is underestimated in the central regions of magnetic spots at different latitudes. Since the Tikhonov regularization applied by our code always smoothes the horizontal gradients, leading to some underestimation of both temperature and magnetic field contrast for small spots, we also calculate a complementary measure of the total magnetic field modulus as well as the total radial, meridional, azimuthal fields by integrating corresponding distributions over the entire stellar surface.

\subsection{Homogeneous temperature distribution}
\label{homtemp}
\begin{table}[!ht]
\centering
\caption{Underestimation of field strength for the inversion with a homogeneous temperature.}
\label{tab2}
\begin{tabular}{cccccc}
\hline\hline
Used Stokes  & Field       & Spot 1                & Spot 2                  & Spot 3              & Spot 4    \\
parameters    & component & lat. 60$^\circ$     &  lat. 40$^\circ$     & lat. 20$^\circ$  & lat. 0$^\circ$  \\
\hline
$V$  & $B_{\rm r}$         & 21\%  & 31\%  & 49\% & 78\% \\
	   & $B_{\rm tot}$      & 21\%  & 31\%  & 40\% & 71\% \\  
\hline	   
$V$  & $B_{\rm m}$        & 42\%  & 69\%  & 58\% & 70\%\\
        & $B_{\rm tot}$      & 42\%  & 68\%  & 52\% & 64\%\\
\hline
$V$   & $B_{\rm a}$         & 30\% & 40\% & 28\% & 56\%  \\
         & $B_{\rm tot}$      & 30\% & 40\% & 28\% & 55\%  \\
\hline 
$IV$  & $B_{\rm r}$   & 10\%  & 12\%  & 12\% & 95\%  \\
        & $B_{\rm tot}$  & 9\%  & 12\%  & 11\% & 22\%  \\
\hline        
$IV$ & $B_{\rm m}$ & 6\%  & 12\%  & 6\% & 20\% \\
        & $B_{\rm tot}$ & 6\%  & 11\%  & 6\% & 20\% \\ 
\hline
$IV$ & $B_{\rm a}$   & 7\% & 8\% & 7\% & 13\% \\
        & $B_{\rm tot}$ & 7\% & 8\% & 7\% & 13\% \\
\hline 
\end{tabular}
\tablefoot{The compared surface elements are taken from the center of each spot.}
\end{table}
\begin{table*}[ht]
\centering
\caption{Integrated field values for the inversions with a homogeneous temperature.}
\label{tab5}
\begin{tabular}{cccc|cc|cc}
\hline\hline
Used Stokes & Field            & True         & Reconstructed  & True         & Reconstructed  & True            & Reconstructed   \\
parameters   & component & map (kG\,rad$^2$) & map (kG\,rad$^2$)          & map (kG\,rad$^2$)  & map (kG\,rad$^2$)          & map (kG\,rad$^2$)    &   map (kG\,rad$^2$) \\
\hline 
$V$               & $B_{\rm r}$           & 2.945 & 2.682  & 0        & 0.340   & 0        & 0.047 \\
                     & $B_{\rm m}$         & 0        & 0.798   & 2.945 & 2.997   & 0        & 0.030 \\                    
                    & $B_{\rm a}$          & 0        & $-0.084$ & 0        & $-0.052$  & 2.945 & 3.616 \\
                     & $B_{\rm tot}$       & 2.945 & 3.574  & 2.945 & 3.550    & 2.945 & 4.117 \\
\hline 
$IV$             &  $B_{\rm r}$          & 2.945 & 3.828  & 0        & $-0.687$ & 0        & 0.265 \\
                     & $B_{\rm m}$         & 0        & 0.497  & 2.945 & 4.065  & 0        & $-0.305$ \\
                     & $B_{\rm a}$          & 0        & $-0.099$ & 0        & $-0.009$ & 2.945 & 3.408 \\
                     & $B_{\rm tot}$       & 2.945 & 5.360  & 2.945  & 4.524 & 2.945 & 4.378 \\
\hline
\end{tabular}
\end{table*}
The maps resulting from the inversions using line profiles simulated for the homogeneous temperature distribution are shown in Figs.~\ref{fig:maps1} and~\ref{fig:maps2}. There are clear differences between the reconstructed magnetic field maps in Fig.~\ref{fig:maps1} and those in Fig.~\ref{fig:maps2} even though the only difference between these inversions was that the code was allowed to vary temperature in the latter case. The most prominent difference between the two sets of maps is the field strength. A closer inspection was made by comparing the field values of specific surface elements of the star in the reconstructed maps to the same elements in the true maps. Four surface elements from the center of each spot were chosen and an average discrepancy value was calculated. These results are reported in Table~\ref{tab2} both for the field modulus and for the specific field component (a large difference between the two values indicates significant crosstalk). The spot positions are however quite accurately reconstructed in both cases even though some spots, especially in Fig.~\ref{fig:maps1}, appear a bit elongated. However, the center of the spots, in terms of the highest magnetic field strength, does not suffer from any shifts in longitude nor in latitude.

The field strengths of individual surface zones are only underestimated by less than 10\% when both Stokes $I$ and $V$ were used for the inversion. When temperature is kept constant the field strength underestimation of individual components varies between 21\% and 78\%. We also integrated each field component and the total field strength over the entire stellar surface. The resulting values can be found in Table~\ref{tab5}.

By inspecting the reconstructed temperature maps shown in Fig.~\ref{fig:maps2} it appears that the areas covered by magnetic spots contain some temperature variations. However, the full range of temperature variation is only about 50--100~K. The temperature spots appear slightly hotter than the rest of the stellar surface.

When temperature is assumed constant the reconstructed radial field spots (see Fig.~\ref{fig:maps1}) are all underestimated in strength. According to Table~\ref{tab2}, the underestimation increases with decreasing latitude. This is also true for the radial spots in Fig.~\ref{fig:maps2}, although the underestimation is not as severe except for the lowest latitude spot which is instead interpreted as a meridional spot by the inversion code. The total field strength of that spot is however only underestimated by 22\%, as can be seen in Table~\ref{tab2}. The total field strength and hence also the radial field strength integrated over the entire stellar surface is 2.945 kG\,rad$^2$ in the true map, see Table~\ref{tab5}. When both magnetic field and temperature are reconstructed, the total field strength and radial field strength are higher, 5.360 kG\,rad$^2$ and 3.828 kG\,rad$^2$ respectively. This can also be seen in Fig.~\ref{fig:maps2} since almost the entire map is covered by a radial field including three spots which are only underestimated with about 10\%. When temperature is kept constant, the total field strength is slightly overestimated while the radial field strength instead is slightly underestimated. The meridional field value, which should be zero, is instead almost 800 G\,rad$^2$, and this can also be seen as crosstalk between radial and meridional fields in Fig.~\ref{fig:maps1}, especially at low latitudes. The two lowest latitude spots has an almost as strong meridional component as the radial one. The increasing crosstalk with decreasing latitude is also evident from the values in Table~\ref{tab2}. The difference in field underestimation between the radial field component and the total field component systematically increases with decreasing latitude.  

The biggest discrepancy between the true and recovered field maps in Figs.~\ref{fig:maps1} and \ref{fig:maps2} can be found for meridional field spots. By comparing the strength of the same spot for each of the three field directions in Table~\ref{tab2} it is evident that the meridional spots are the weakest for three out of the four spots when only Stokes $V$ is used for inversions. The underestimation of the meridional field component varies between 49\% and 70\% in this case. The situation is almost the opposite when both Stokes $I$ and $V$ are used for inversions since now two of the meridional spots are the strongest since they are only underestimated by 6--20\%. The integrated total and meridional field values are however overestimated in both reconstructions. Crosstalk can also be seen and once again it is strongest between meridional and radial components at low latitudes. The overall structure of the spots in the meridional maps in Figs.~\ref{fig:maps1} and \ref{fig:maps2} are somewhat similar in the sense that the two high latitude spots are connected, and the two low latitude spots are spread out towards lower latitudes. Even so, the spots in Fig.~\ref{fig:maps2} are better defined since there is a larger difference in field strength between the actual spots and the field outside the spots.

The reconstructed azimuthal magnetic field spots suffer least from crosstalk but are however quite similar to the meridional field maps in terms of structure. The two high-latitude spots are overlapping while the two low-latitude spots spread out towards lower latitudes. The spots in Fig.~\ref{fig:maps2} are still more clearly defined than for the corresponding inversion using Stokes $V$ alone. Once again the integrated field strengths of the reconstructed maps exceed those of the true map. The field strength is somewhere in between the values for the radial and meridional spots at high latitudes, but higher at low latitudes.   

\begin{table}[!b]
\centering
\caption{Underestimation of field strength for the inversions with an inhomogeneous temperature distribution.}
\label{tab3}
\begin{tabular}{cccccc}
\hline\hline
Used Stokes  & Field   & Spot 1                & Spot 2                  & Spot 3              & Spot 4 \\
parameters    & component & lat. 60$^\circ$ &  lat. 40$^\circ$     & lat. 20$^\circ$  & lat. 0$^\circ$ \\
\hline
$V$  & $B_{\rm r}$   & 82\%  & 85\%  & 89\% & 94\%  \\
        & $B_{\rm tot}$ & 82\%  & 85\%  & 87\% & 92\%  \\
\hline        
$V$  & $B_{\rm m}$ & 87\%  & 95\%  & 94\% & 94\% \\
        & $B_{\rm tot}$ & 86\%  & 94\%  & 90\% & 91\% \\
\hline       
$V$    & $B_{\rm a}$   & 83\% & 85\% & 85\% & 91\% \\
         & $B_{\rm tot}$ & 83\% & 85\% & 85\% & 91\% \\
\hline 
$IV$  & $B_{\rm r}$   & 46\%  & 45\%  & 60\% & 76\%  \\
        & $B_{\rm tot}$  & 45\%  & 45\%  & 55\% & 67\%  \\
\hline      
$IV$ & $B_{\rm m}$ & 70\%  & 84\%  & 70\% & 73\% \\
        & $B_{\rm tot}$ & 69\%  & 81\%  & 64\% & 67\% \\
\hline       
$IV$ & $B_{\rm a}$   & 46\% & 53\% & 47\% & 60\% \\
        & $B_{\rm tot}$ & 46\% & 53\% & 47\% & 60\% \\
\hline 
\end{tabular}
\tablefoot{The compared surface elements are taken from the center of each spot.}
\end{table}

\subsection{Inhomogeneous temperature distribution}
\label{inhomtemp}

\begin{table*}[!ht]
\centering
\caption{Integrated field values for the inversions with an inhomogeneous temperature distribution.}
\label{tab6}
\begin{tabular}{cccc|cc|cc}
\hline\hline
Used Stokes & Field            & True         & Reconstructed  & True         & Reconstructed  & True            & Reconstructed   \\
parameters   & component & map (kG\,rad$^2$) & map (kG\,rad$^2$)          & map (kG\,rad$^2$)  & map (kG\,rad$^2$)          & map (kG\,rad$^2$)    &   map (kG\,rad$^2$) \\
\hline 
$V$               & $B_{\rm r}$           & 2.945 & 1.243  & 0        & 0.235   & 0        & 0.072 \\
                     & $B_{\rm m}$         & 0        & 0.423   & 2.945 & 0.553   & 0        & 0.003 \\
                     & $B_{\rm a}$          & 0        & $-0.090$ & 0        & $-0.016$  & 2.945 & 1.313 \\
                     & $B_{\rm tot}$  & 2.945 & 1.479  & 2.945 & 0.864    & 2.945 & 1.391 \\
\hline 
$IV$             &  $B_{\rm r}$          & 2.945 & 2.159  & 0        & 0.647 & 0        & $-0.091$ \\
                     & $B_{\rm m}$         & 0        & 0.474  & 2.945 & 1.645 & 0        & 0.184 \\
                     & $B_{\rm a}$          & 0        & $-0.078$ & 0        & $-0.214$ & 2.945 & 2.233 \\
                     & $B_{\rm tot}$  & 2.945 & 2.994  & 2.945  & 2.539 & 2.945 & 2.721 \\
\hline
\end{tabular}
\end{table*} 

The results of the inversions for magnetic features coinciding with low-temperature spots are presented in Figs.~\ref{fig:maps3} and~\ref{fig:maps4}. As can be seen from the reconstructed maps shown in Fig.~\ref{fig:maps3}, the magnetic field strength is strongly underestimated for all three field components when temperature variations are ignored. By comparing the field values of specific surface elements in the same way as in the first experiment, we inferred that the field strengths are underestimated by as much as 80--95\% (see Table~\ref{tab3}). On the other hand, the temperature reconstruction in the inversion using both Stokes $I$ and $V$ once again proved to be very accurate. There is a tendency for a slight overestimation, but only by approximately 250~K. 
\begin{figure*}[!ht]
\centering
\subfigure{\includegraphics[scale=0.55,angle=90]{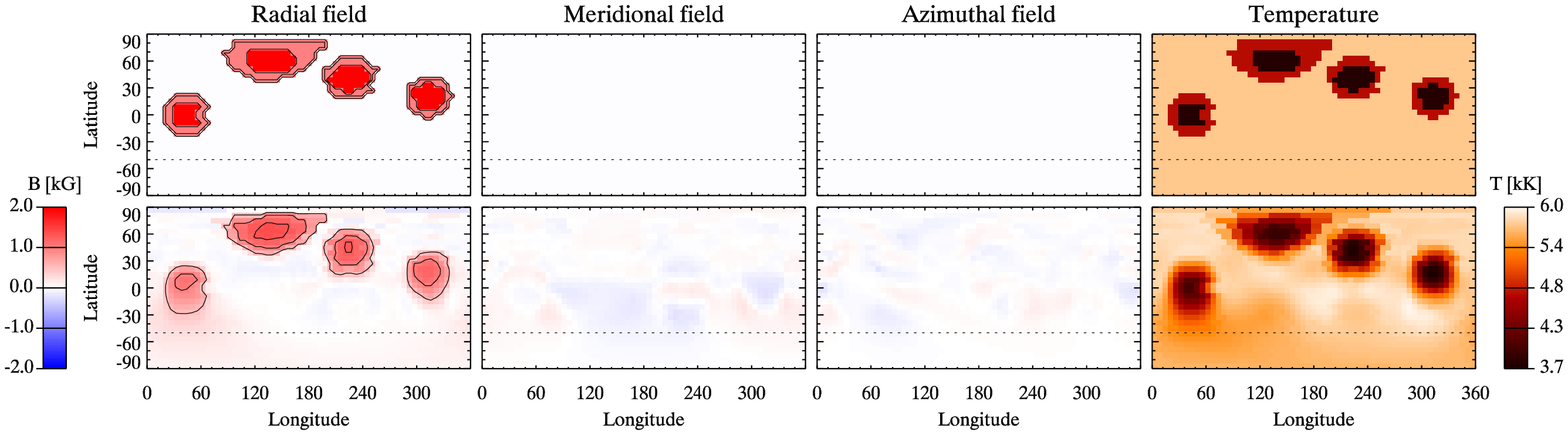}
\label{fig:subfig21}}
\centering
\subfigure{\includegraphics[scale=0.55,angle=90]{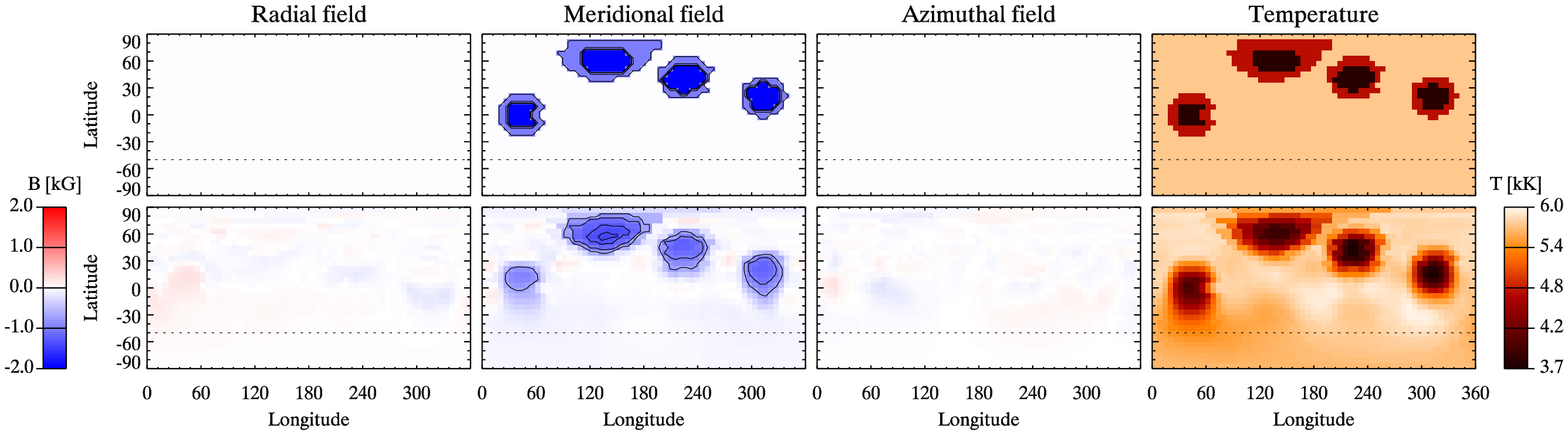}
\label{fig:subfig41}}
\centering
\subfigure{\includegraphics[scale=0.55,angle=90]{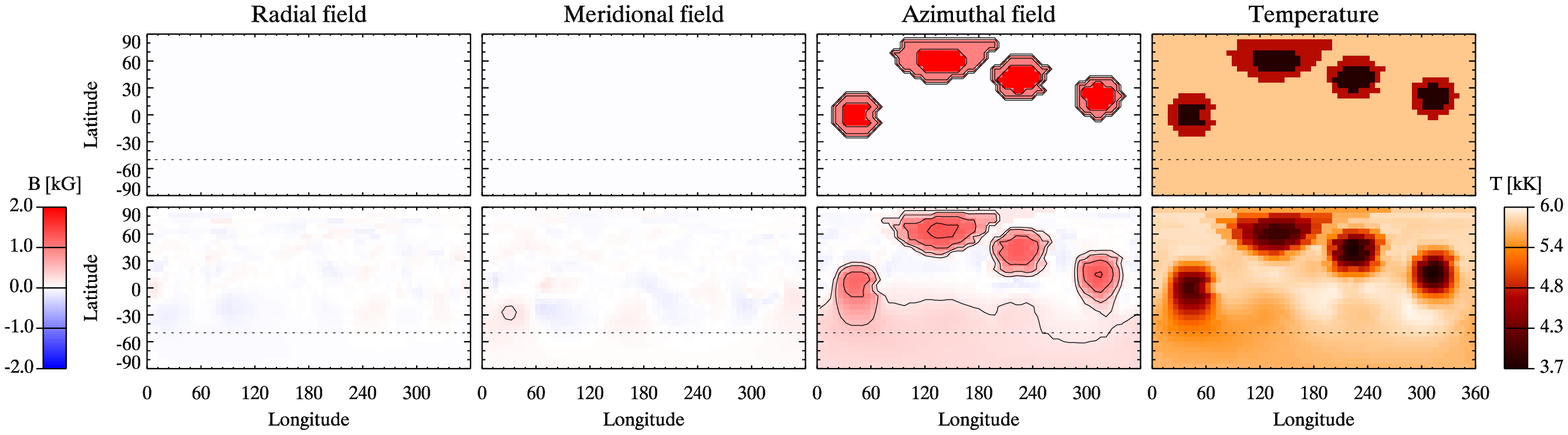}
\label{fig:subfig61}}
\caption[Optional caption for list of figures]{Same as for Fig.~\ref{fig:maps4} but here all four Stokes parameters were used in self-consistent ZDI inversions. }
\label{fig:maps5}
\end{figure*}
As in the first experiment, the field strengths of the reconstructed radial spots shown in Fig.~\ref{fig:maps3} seem to decrease with decreasing latitude. When temperature variations are taken into account (Fig.~\ref{fig:maps4}) the two high-latitude spots have similar strengths while the two low-latitude spots are distorted similarly to the first experiment. Once again there is crosstalk primarily between the radial and meridional field components at low latitudes in both cases. This can be seen directly in the reconstructed maps, but also in Table~\ref{tab3} since once again the underestimation of the total field strength is less compared to the underestimation of the radial field strength at low latitudes. Otherwise there are not many similarities between the two radial field maps. The underestimation of the radial field strengths when only Stokes $V$ is used for inversion varies between 82\% and 94\%. The underestimation is not as severe when temperature is taken into account and varies between 45\% and 76\% instead. The integrated field values, which can be found in Table~\ref{tab6}, show that the radial field component and the total field strength are almost twice as large when Stokes $I$ and $V$ are used for inversions compared to when only Stokes $V$ is used. Despite this, the meridional field strengths, which should be zero when the true spots are purely radial, are comparable in strength, and so are the azimuthal field strengths. In other words, even though the radial field strength is so much weaker when temperature variations are ignored, there is still almost as much crosstalk as when temperature is taken into account. The spot geometry is quite accurately reproduced in Fig.~\ref{fig:maps4} (self-consistent magnetic and temperature inversion from both Stokes $I$ and $V$) while it is almost impossible to relate to the original maps in Fig.~\ref{fig:maps3} (magnetic mapping from Stokes $V$ alone).

\begin{table}[!b]
\centering
\caption{Underestimation of field strength for the self-consistent inversions using all four Stokes parameters.}
\label{tab4}
\begin{tabular}{cccccc}
\hline\hline
Used Stokes  & Field   & Spot 1                & Spot 2                  & Spot 3              & Spot 4 \\
parameters    & component & lat. 60$^\circ$ &  lat. 40$^\circ$     & lat. 20$^\circ$  & lat. 0$^\circ$ \\
\hline
$IQUV$  & $B_{\rm r}$      & 33\%  & 36\%  & 41\% & 49\%  \\
               & $B_{\rm tot}$   & 32\%  & 36\%  & 40\% & 49\%  \\
\hline               
$IQUV$   & $B_{\rm m}$   & 31\%  & 44\%  & 41\% & 62\% \\
               & $B_{\rm tot}$  & 31\%  & 44\%  & 41\% & 61\% \\
\hline             
$IQUV$   & $B_{\rm a}$   & 34\% & 37\% & 36\% & 48\% \\
               & $B_{\rm tot}$   & 34\% & 37\% & 36\% & 43\% \\
\hline 
\end{tabular}
\tablefoot{The compared surface elements are taken from the center of each spot.}
\end{table}

\begin{table*}[!ht]
\centering
\caption{Integrated field values when all four Stokes parameters are used for inversions.}
\label{tab7}
\begin{tabular}{cccc|cc|cc}
\hline\hline
Used Stokes & Field            & True         & Reconstructed  & True         & Reconstructed  & True            & Reconstructed   \\
parameters   & component & map (kG\,rad$^2$) & map (kG\,rad$^2$)          & map (kG\,rad$^2$)  & map (kG\,rad$^2$)          & map (kG\,rad$^2$)    &   map (kG\,rad$^2$) \\
\hline 
$IQUV$         & $B_{\rm r}$           & 2.945 & 2.480  & 0        & 0.099   & 0        & $-0.017$ \\
                     & $B_{\rm m}$         & 0        & 0.054   & 2.945 & 2.363   & 0        & $-0.096$ \\
                     & $B_{\rm a}$          & 0        & $-0.001$ & 0        & 0.019  & 2.945 & 3.029 \\
                     & $B_{\rm tot}$  & 2.945 & 3.157  & 2.945 & 2.802    & 2.945 & 3.458 \\
\hline 
\end{tabular}
\end{table*}

Comparing the quality of all the reconstructions in all the experiments, the worst one can be found in the experiment for the meridional spots when temperature variations are not taken into account. The meridional field strength is underestimated by as much as 87\% for one spot and by 94--95\% for the other three spots. The situation is somewhat improved when temperature is taken into account since the field is underestimated by 70--84\% instead. The integrated total and meridional field strengths are almost three times higher when inversions are made using both Stokes $I$ and $V$ compared to when only Stokes $V$ is used. However, there is still a clear crosstalk between the meridional and radial field components (see Fig.~\ref{fig:maps4}). 

The reconstruction of the azimuthal field spots follows the same pattern as for the other two field directions in the sense that recovery of the spot strength and geometry are significantly improved when temperature variations are taken into account. The azimuthal field strengths are underestimated by 46--60\% compared to 83--91\% for the inversion neglecting temperature spots. The integrated total azimuthal field strength is twice as large. Similar to the first experiment, the azimuthal spots are most robust and suffer very little from crosstalk. Furthermore, the strengths of the two lowest latitude spots are higher compared to the corresponding radial and meridional spots.     

\subsection{All four Stokes parameters}
\label{4stokes}

The maps obtained in the inversions using all four Stokes parameters are shown in Fig.~\ref{fig:maps5}. The field strength is underestimated by 31\% to 62\% (see Table~\ref{tab4}), but the spot geometry is very accurately reconstructed, with the only exception being the lowest latitude azimuthal spot. Almost none of the crosstalk seen in the first two experiments is visible in the full Stokes vector inversion. The temperature is once again correctly reproduced to within 200~K in general. 

As in the other two experiments, the radial field strength underestimation is systematically increasing from 33\% for the highest-latitude spot to 49\% for the lowest-latitude spot. The total field strength is higher than the true value, but the radial field strength is lower, see Table~\ref{tab7}. The spot geometry and positions are accurately recovered and the map shows little spurious magnetic flux outside the spots. No significant crosstalk is detected and it appears that the ability of the inversion code to distinguish the radial and meridional component is now as good as it was for the radial and azimuthal components in the inversions with circular polarization alone.

The geometry of meridional spots is correctly recovered and their strengths are underestimated by 31--62\%. Here the integrated total field and meridional field values are slightly underestimated.

The azimuthal spots have similar strengths to the radial spots since they are underestimated by 34--48\%. The lowest-latitude spot does however seem to spread over lower latitudes making the reconstructed spot topology less well-defined compared to the other two maps. This can also be seen in the integrated field values, which are slightly overestimated even though the field strengths of the centers of spots are underestimated. Once again, the reconstructed field strengths inside the two low-latitude spots are higher for the azimuthal field spots compared to the same experiments with radial and meridional field spots.

\section{Discussion}
\label{dis}

In this study we carried out numerical ZDI simulations by inverting spectra generated using a standard grid of stellar model atmospheres. In other words, following numerous previous conventional temperature DI studies, we have assumed that the thermal structure of a starspot can be approximated by a simple radiative equilibrium model atmosphere. Studies of the Sun demonstrated validity of such an approximation at least for the sunspot interiors \citep[see][and references therein]{1998A&A...329..747S,2003A&ARv..11..153S}. Until realistic magnetic starspot models become widely available, this is the only feasible approach for the model-atmosphere-based active-star DI. In any case, we are mostly concerned with including in the magnetic mapping a dominant effect of the intensity contrast due to presence of hot and cool surface features. The temperature dependence of the local Stokes $I$ profiles is the second most important factor, properly incorporated in our calculations. On top of that, possible deviations of the $T$\,-\,$\tau$ relation from the predictions of hydrostatic model atmospheres might induce line shape changes of Stokes parameters. But, even if present, this effect is going to be by far less important than the scaling of the polarization profiles by the local continuum intensity and by the factor reflecting temperature sensitivity of specific spectral lines.

One of the unexpected outcomes of our numerical experiments is the difference of magnetic maps inferred from Stokes $IV$ and from $V$ alone in the absence of temperature spots. When the temperature is homogeneous one might expect that the resulting magnetic field maps should not depend on whether temperature was reconstructed since in both cases the code ends with a nearly identical temperature distribution lacking significant inhomogeneities. But the magnetic maps presented in Figs.~\ref{fig:maps1} and~\ref{fig:maps2} are not identical although the temperature is the same to within 50--100 K. The explanation to this discrepancy might be that information about the magnetic field is also contained in Stokes $I$ which is made available and taken into account when Stokes $I$ is used for simultaneous temperature mapping. In order to test this further, we performed another inverse calculation of the stellar surface with magnetic spots and a homogeneous temperature using both Stokes $I$ and $V$, but excluding the Zeeman effect in the forward calculations of Stokes $I$. The magnetic maps obtained in this case resemble more closely the magnetic distributions shown in Fig.~\ref{fig:maps1}, which were reconstructed using only Stokes $V$, rather than the ones illustrated in Fig.~\ref{fig:maps2}.  

\begin{figure}[!h]
\centering
\includegraphics[scale=0.5,angle=90]{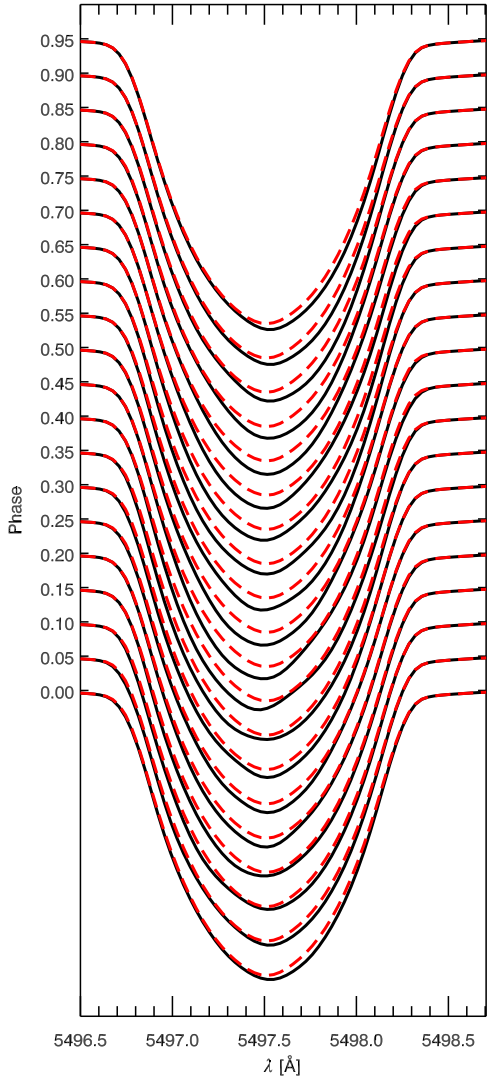}
\caption{Stokes $I$ profiles from a stellar surface with homogeneous temperature. The black solid lines represents the Stokes $I$ profiles when the stellar surface contained magnetic spots and the red dashed lines represents Stokes $I$ without any magnetic spots.}
\label{stokesi}
\end{figure} 

\begin{figure}[!ht]
\centering
\includegraphics[scale=0.5,angle=90]{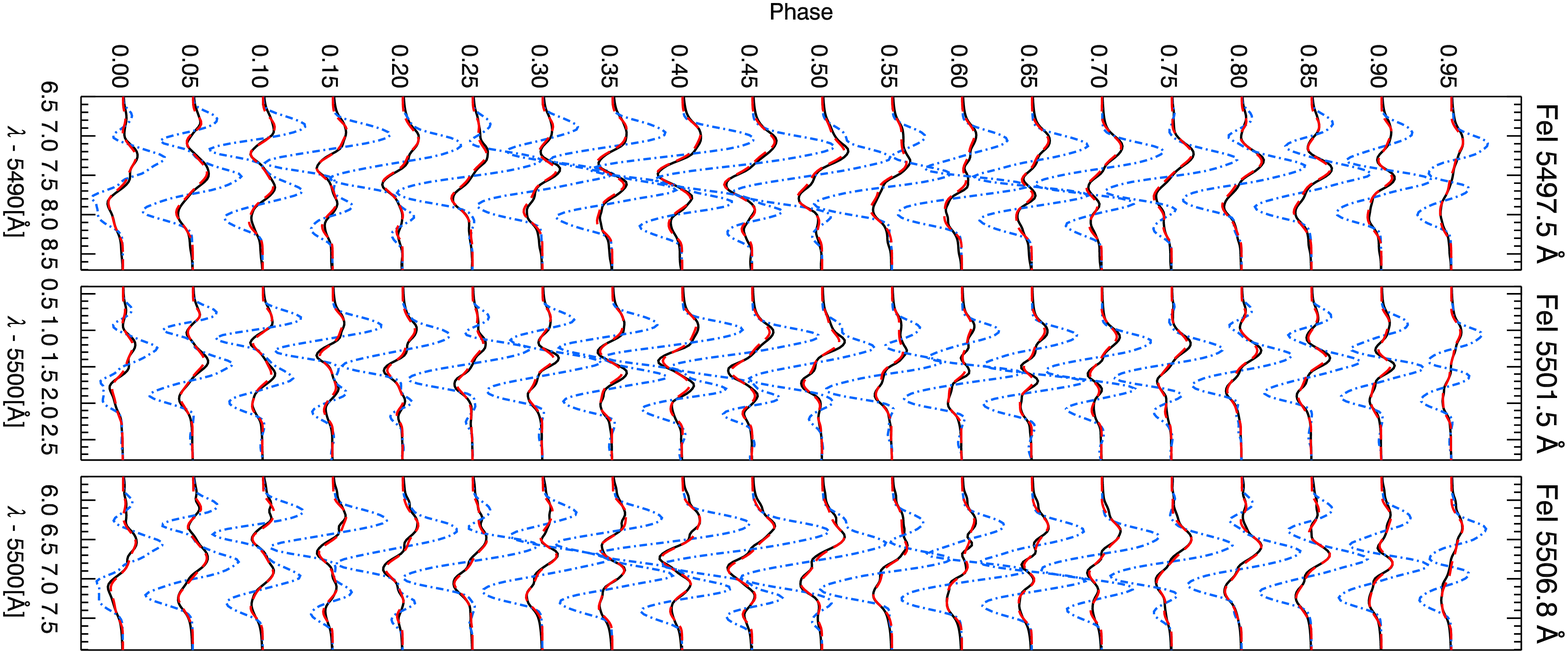}
\caption{Line profiles from forward and inverse calculations. The black and red lines correspond to the line profiles from forward and inverse calculations respectively of a stellar surface with magnetic spots and cool temperature spots. The blue lines represents line profiles from forward calculations of a stellar surface with magnetic spots and a homogeneous temperature distribution.}
\label{fig:stokes}
\end{figure}

We found that for our assumed magnetic spot configuration Stokes $I$ is indeed affected by the magnetic field. The Zeeman effect leads to splitting and broadening of local spectral lines and this shows up as a broadening and deepening of the corresponding disk-integrated Stokes $I$ profile. In Fig.~\ref{stokesi} we have plotted all Stokes $I$ profiles from the first experiment, i.e. when the stellar surface had magnetic spots but a homogeneous temperature together with another set of Stokes $I$ profiles of a stellar surface with the same homogeneous temperature but without any magnetic spots. There is a clear difference between the two sets of profiles since the one where magnetic spots were included is rotationally modulated. Thus, not only does temperature variations have to be taken into account when magnetic mapping is done, but a magnetic field also have to be taken into account when doing temperature mapping. That is certainly one of the reasons to why the temperature reconstruction is accurate in our experiments. If only temperature is reconstructed using Stokes $I$ in the case with magnetic spots on a surface with homogeneous temperature, the spurious temperature variations reach to about 500~K locally. This should be compared to a difference of 100~K when Stokes $V$ is included for simultaneous temperature and magnetic field reconstruction as in the first experiment. 

We also performed the same test for the surface distribution used in the second experiment, i.e. the map with cool magnetic spots. When only Stokes $I$ was used to reconstruct the temperature map ignoring magnetic field, the temperature variations were found to be close to the temperature distribution reconstructed self-consistently using both Stokes $I$ and $V$. This is, perhaps, not so unexpected since the influence of the Zeeman effect on Stokes $I$ is significantly smaller than the distortions caused by the cool temperature spots. These results imply that the conventional temperature DI, which does not take the Zeeman effect into account, is not necessarily incorrect, at least for the magnetic field strengths comparable to those investigated in our study.

Perhaps it is not so surprising that the magnetic field reconstruction is so poor when temperature variations are ignored. The amplitude of the Stokes $V$ line profiles depends on both temperature and magnetic field strength. This has also been discussed before, for example by \citet{Donati1997}. Instead of simultaneously reconstructing temperature, they are using a field strength of 500 G which they say corresponds to a 2 kG field, in Stokes $V$, with 50\% filling factors and 50\% temperature-induced signal dilution. When temperature is not reconstructed simultaneously with magnetic field, the temperature of the magnetic spots is assumed to be the same as the photospheric temperature. The low amplitude of the polarized line profile is then incorrectly interpreted as if the magnetic field strength is low. When temperature is taken into account, as in our magnetic inversion technique, the code adjusts the field strength according to the appropriate local temperature. The low amplitude of the polarized profile is therefore not misinterpreted in terms of low magnetic field strength. To illustrate the striking difference in amplitude of magnetic signal associated with cool spots compared to magnetic spots at photospheric temperature, we show the respective Stokes $V$ profiles in Fig.~\ref{fig:stokes}. 

Furthermore, besides a general major underestimate of the magnetic flux in the ZDI without accounting for temperature spots, our experiments suggest that magnetic field topology also cannot be correctly reconstructed with such an inconsistent approach. Comparison of Figs.~\ref{fig:maps3} and \ref{fig:maps4} clearly shows that magnetic reconstruction ignoring temperature spots does not yield a faint copy of the correct map. Instead, we get highly distorted magnetic configurations, which in many cases bear little resemblance to the input magnetic maps. Thus, contrary to the assumption of the ZDI studies which treat Stokes $I$ and $V$ fully independently \citep[e.g.][]{Donati1997,Donati1999}, the lack of self-consistency between interpretation of Stokes $I$ and $V$ actually jeopardizes all aspects of magnetic mapping.

\begin{figure}[!t]
\centering
\subfigure[Line profiles when temperature is homogeneous and only Stokes $V$ is used for inversions.]{\includegraphics[scale=0.35,angle=90]{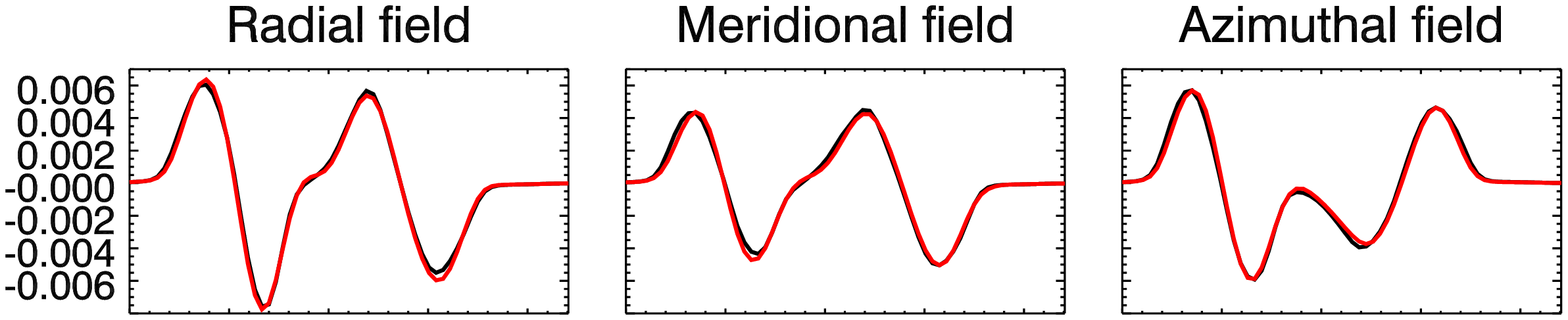}
\label{fig:3prof1}}
\subfigure[Line profiles when temperature is homogeneous and both Stokes $I$ and $V$ are used for inversions.]{\includegraphics[scale=0.35,angle=90]{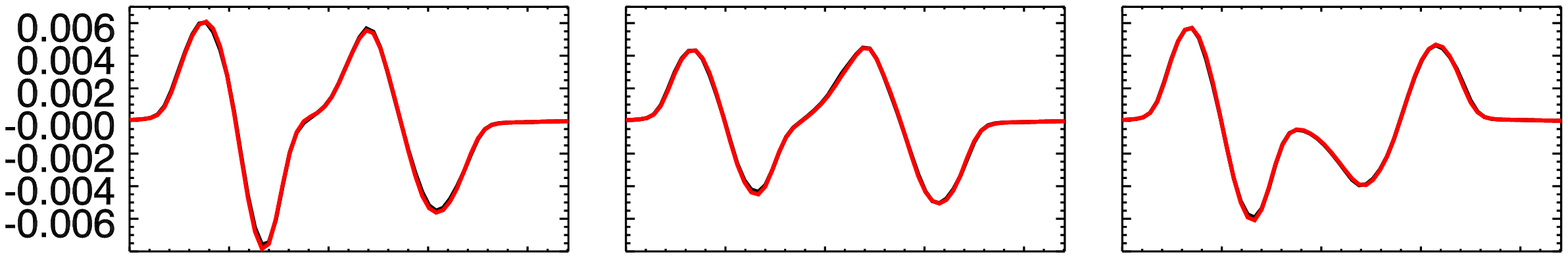}
\label{fig:3prof2}}
\subfigure[Line profiles when temperature is inhomogeneous and only Stokes $V$ is used for inversions.]{\includegraphics[scale=0.35,angle=90]{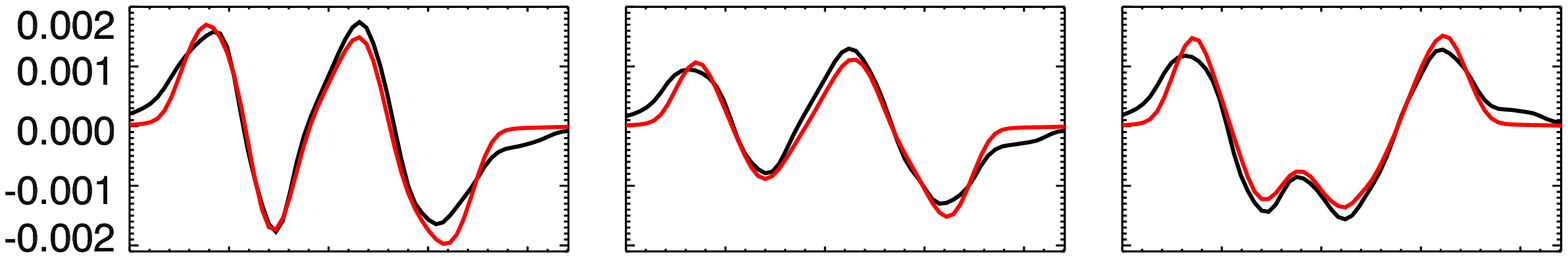}
\label{fig:3prof3}}
\subfigure[Line profiles when temperature is inhomogeneous and both Stokes $I$ and $V$ are used for inversions.]{\includegraphics[scale=0.35,angle=90]{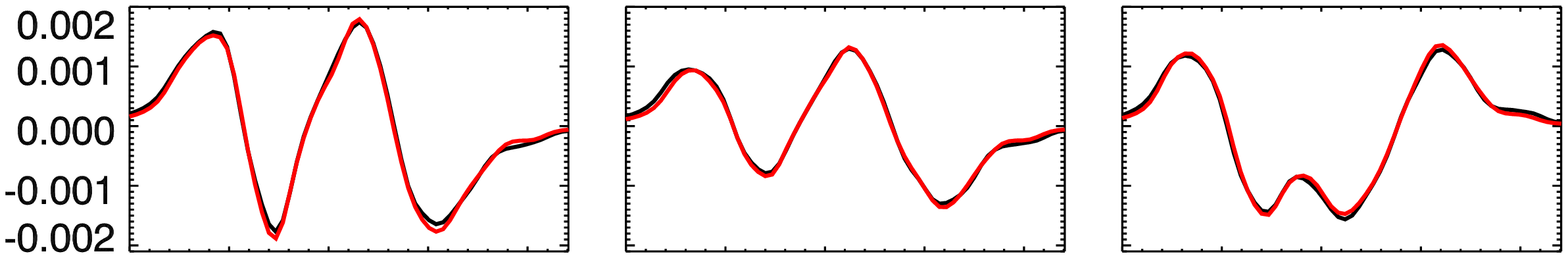}
\label{fig:3prof4}}
\subfigure[Line profiles when temperature is inhomogeneous and all four Stokes parameters are used for inversions.]{\includegraphics[scale=0.35,angle=90]{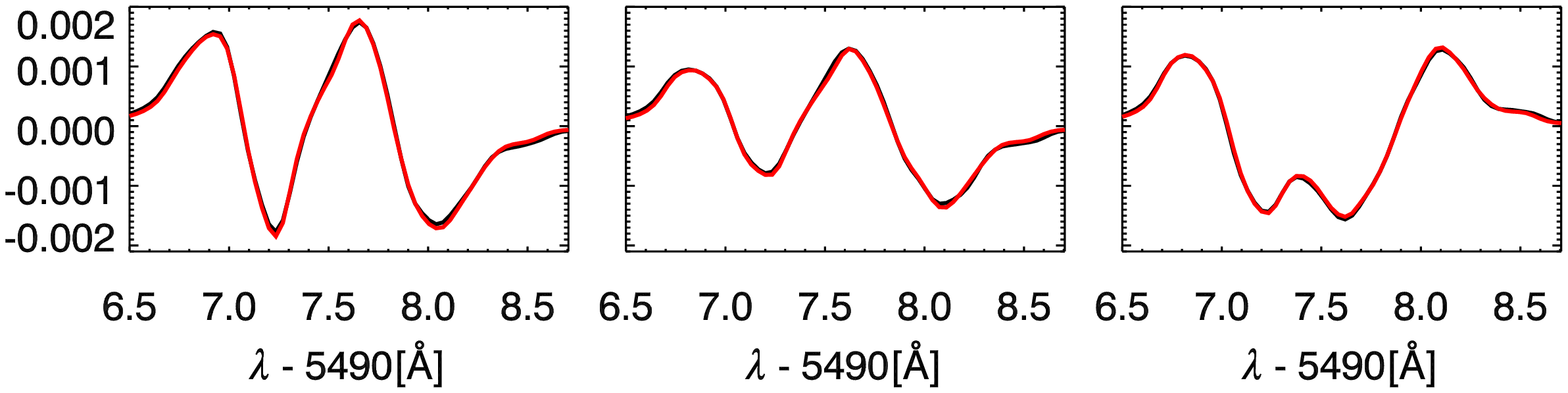}
\label{fig:3prof5}}
\caption[Optional caption for list of figures]{
Example of Stokes $V$ profile fits for the rotational phase 0.1 for each of the inversion tests with radial field. Black lines represent the true line profiles produced with forward calculation and the red lines show corresponding fits by \inv\ code.
}
\label{fig:3prof}
\end{figure}

We have verified that these inversion results are not due to a large difference in the quality of profile fits for different inversions. Fig.~\ref{fig:3prof}  shows the typical fits to line profiles corresponding to different experiments. Evidently, the difference is not that large. The worst fits are, not surprisingly, found for fixed-temperature fitting of the Stokes $V$ profiles associated with an inhomogeneous temperature distribution. In this case, a single-component atmosphere assumed by the code is physically incorrect and cannot reproduce all details of the circular polarization profile by adjusting the field strength and orientation. But even in this case the fits are not grossly incorrect in the sense that the profile amplitudes are not severely underestimated. 
What can also be seen from Fig.~\ref{fig:3prof} is that the quality of the fit does not significantly depend on the field direction.      

We found that even in the simultaneous inversion using both Stokes $I$ and $V$ magnetic reconstruction suffers a number of crosstalks. In particular, the crosstalk between the radial and meridional field components is prominent and can be explained by the similarity in the corresponding polarized line profiles \citep{Donati1997}. Circularly polarized light is diagnosing the line-of-sight component of the magnetic field vector. A magnetic spot with a meridional field will therefore produce an anti-symmetric Stokes $V$ signature, quite similar to that of a radial field spot, since the projection of the magnetic field vector on the line-of-sight will not change sign with rotation phase for a given spot. An azimuthal field will, however, produce a different dynamical Stokes $V$ signature since the projected magnetic field vector will change sign in the center of the stellar disk as the spot moves relative to the observer due to stellar rotation. In Fig.~\ref{fig:3prof} Stokes $V$ line profiles taken from the same rotational phase can be seen. The profiles corresponding to radial and meridional field spots are similar in shape while the profiles representing azimuthal field spots are very different compared to the other two. 

The crosstalk between radial and meridional fields is strongest at low latitudes. This feature is easy to explain because in this case magnetic spots will be observed from a limited range of viewing angles. In general, the magnitude of the line-of-sight projection of the magnetic field vector for a spot with a constant radial field will change with phase while the magnitude of a projected field vector for a spot with a constant meridional field will be almost constant with rotational phase. This allows the code to distinguish radial and meridional fields at higher latitudes which are observable over a larger range of phases. However, low-latitude spots are seen from approximately the same aspect angle, resulting in approximately constant projected field vector. This makes it difficult to distinguish radial and meridional fields.

The meridional field is the most difficult field orientation to reproduce. On the one hand, as already mentioned, a meridional field might be interpreted as a radial field because of the similarity in the respective Stokes $V$ profile shapes. On the other hand, a field vector of a radial or an azimuthal field will at some rotational phase point more or less straight towards the observer, depending on the stellar inclination and phase of course, making the projected field vector almost as large as the local field vector. A meridional field vector will, however, always be almost perpendicular to the line-of-sight, making the line-of-sight magnetic vector projection small and hence resulting in a small amplitude of the circular polarization profile. These aspects conspire to make recovery of the meridional field very challenging.

When all four Stokes parameters are used, most of the crosstalk is removed. By including linear polarization, which is sensitive to the transverse component of the magnetic field vector, a lot more information about the orientation of the field can be retrieved, as already demonstrated by the ZDI experiments for magnetic Ap stars \citep{Kochukhov02}. By comparing Fig.~\ref{fig:maps3} to Fig.~\ref{fig:maps5}, which have the same spot set-up, one can see that the reconstructed fields are generally stronger when all four Stokes parameters are used, especially for the meridional field. Even though the line-of-sight component of a meridional field vector is small, the perpendicular component is large and this is detected by the linear polarization.

\section{Conclusions}
\label{con}

We have carried out a series of numerical experiments with Zeeman Doppler imaging, aiming to explore reliability of reconstruction of different magnetic field geometries and testing to what extent independent reconstruction of temperature and magnetic spots is meaningful for stars with high-contrast temperature inhomogeneities. We performed these tests under ideal conditions, adopting optimal stellar parameters in terms of $v_{\rm e}\sin i$ and $i$, using an excellent phase coverage and adding no noise to the simulated line profiles.

The main conclusions of our study are the following:
\begin{itemize}
\item
For stars with significant temperature inhomogeneities a reliable magnetic field reconstruction with ZDI is essentially impossible when temperature is not taken into account during mapping of magnetic spots. If a multi-component nature of the active-star surface is not taken into account in magnetic mapping, the magnetic field strength inside cool spots is grossly underestimated and the overall geometry of the field is not recovered correctly.

\item
Our tests also showed that even if the temperature distribution is homogeneous there will still be a discrepancy between reconstructed magnetic field maps depending on if Stokes $I$ is used in mapping or not. The temperature reconstruction itself was accurate to within 100~K and should therefore not have caused the relatively large difference in the reconstructed magnetic field maps. Instead, we found that additional information contained in the Zeeman broadening and intensification of Stokes $I$ profiles helps to constrain magnetic maps.

\item
The shape of Stokes $I$ line profiles is affected by a magnetic field making them broader and rotationally modulated  even in the absence of temperature inhomogeneities. It may therefore be important to take a magnetic field into account during temperature mapping of active cool stars with especially strong fields.

\item
According to our results, simultaneous recovery of magnetic field and temperature is feasible with a self-consistent interpretation of the Stokes $I$ and $V$ profiles of individual spectral lines. We did not find significant crosstalks between temperature and magnetic maps. Instead, reconstruction of magnetic field is greatly improved when appropriate local temperature variations are taken into account. Thus, there are no reasons whatsoever to keep temperature and magnetic inversions separate, as done by the majority of previous and current ZDI studies.

\item
Magnetic mapping using Stokes $I$ and $V$ suffers from a number of systematic artifacts. For example, it is difficult to distinguish radial and meridional fields at low latitudes and the strength of the reconstructed radial field generally decreases with decreasing latitude. 
At the same time, azimuthal field is reliably recovered for a wide range of spot latitudes and does not suffer from a crosstalk with either radial or meridional field.

\item
When the linear polarization data are incorporated in the ZDI inversions, the reconstructed field topology becomes noticeably more accurate compared to the inversions with partial Stokes data sets. In particular, by including all four Stokes parameters one can avoid crosstalks between different field components.
\end{itemize}

\begin{acknowledgements}
OK is a Royal Swedish Academy of Sciences Research Fellow, supported by grants from Knut and Alice Wallenberg Foundation and Swedish Research Council.
\end{acknowledgements}

\bibliographystyle{aa}
\bibliography{astro_ref_v1}

\begin{thebibliography}{23}
\expandafter\ifx\csname natexlab\endcsname\relax\def\natexlab#1{#1}\fi

\bibitem[{{Brown} {et~al.}(1991){Brown}, {Donati}, {Rees}, \&
  {Semel}}]{Brown91}
{Brown}, S.~F., {Donati}, J.-F., {Rees}, D.~E., \& {Semel}, M. 1991, \aap, 250,
  463

\bibitem[{{Catala} {et~al.}(2007){Catala}, {Donati}, {Shkolnik}, {Bohlender},
  \& {Alecian}}]{Catala2007}
{Catala}, C., {Donati}, J.-F., {Shkolnik}, E., {Bohlender}, D., \& {Alecian},
  E. 2007, \mnras, 374, L42

\bibitem[{{Donati}(1999)}]{Donati1999}
{Donati}, J.-F. 1999, \mnras, 302, 457

\bibitem[{{Donati} \& {Brown}(1997)}]{Donati1997}
{Donati}, J.-F. \& {Brown}, S.~F. 1997, \aap, 326, 1135

\bibitem[{{Donati} {et~al.}(2003){Donati}, {Collier Cameron}, {Semel},
  {Hussain}, {Petit}, {Carter}, {Marsden}, {Mengel}, {L{\'o}pez Ariste},
  {Jeffers}, \& {Rees}}]{Donati2003}
{Donati}, J.-F., {Collier Cameron}, A., {Semel}, M., {et~al.} 2003, \mnras,
  345, 1145

\bibitem[{{Donati} {et~al.}(2006){Donati}, {Howarth}, {Jardine}, {Petit},
  {Catala}, {Landstreet}, {Bouret}, {Alecian}, {Barnes}, {Forveille},
  {Paletou}, \& {Manset}}]{Donati06}
{Donati}, J.-F., {Howarth}, I.~D., {Jardine}, M.~M., {et~al.} 2006, \mnras,
  370, 629

\bibitem[{{Gustafsson} {et~al.}(2008){Gustafsson}, {Edvardsson}, {Eriksson},
  {J{\o}rgensen}, {Nordlund}, \& {Plez}}]{Gustafsson2008}
{Gustafsson}, B., {Edvardsson}, B., {Eriksson}, K., {et~al.} 2008, \aap, 486,
  951

\bibitem[{{Kochukhov} {et~al.}(2004){Kochukhov}, {Drake}, {Piskunov}, \& {de la
  Reza}}]{Kochukhov2004}
{Kochukhov}, O., {Drake}, N.~A., {Piskunov}, N., \& {de la Reza}, R. 2004,
  \aap, 424, 935

\bibitem[{{Kochukhov} \& {Piskunov}(2002)}]{Kochukhov02}
{Kochukhov}, O. \& {Piskunov}, N. 2002, \aap, 388, 868

\bibitem[{{Kochukhov} \& {Wade}(2010)}]{Kochukhov10}
{Kochukhov}, O. \& {Wade}, G.~A. 2010, \aap, 513, A13

\bibitem[{{Kochukhov} {et~al.}(2012){Kochukhov}, {Wade}, \&
  {Shulyak}}]{Kochukhov12}
{Kochukhov}, O., {Wade}, G.~A., \& {Shulyak}, D. 2012, \mnras, 421, 3004

\bibitem[{{Kupka} {et~al.}(1999){Kupka}, {Piskunov}, {Ryabchikova}, {Stempels},
  \& {Weiss}}]{Kupka1999}
{Kupka}, F., {Piskunov}, N., {Ryabchikova}, T.~A., {Stempels}, H.~C., \&
  {Weiss}, W.~W. 1999, \aaps, 138, 119

\bibitem[{{L{\"u}ftinger} {et~al.}(2010){L{\"u}ftinger}, {Kochukhov},
  {Ryabchikova}, {Piskunov}, {Weiss}, \& {Ilyin}}]{Luftinger10}
{L{\"u}ftinger}, T., {Kochukhov}, O., {Ryabchikova}, T., {et~al.} 2010, \aap,
  509, A71

\bibitem[{{Petit} {et~al.}(2008){Petit}, {Dintrans}, {Solanki}, {Donati},
  {Auri{\`e}re}, {Ligni{\`e}res}, {Morin}, {Paletou}, {Ramirez Velez},
  {Catala}, \& {Fares}}]{Petit2008}
{Petit}, P., {Dintrans}, B., {Solanki}, S.~K., {et~al.} 2008, \mnras, 388, 80

\bibitem[{{Petit} {et~al.}(2004){Petit}, {Donati}, {Wade}, {Landstreet},
  {Bagnulo}, {L{\"u}ftinger}, {Sigut}, {Shorlin}, {Strasser}, {Auri{\`e}re}, \&
  {Oliveira}}]{Petit2004}
{Petit}, P., {Donati}, J., {Wade}, G.~A., {et~al.} 2004, \mnras, 348, 1175

\bibitem[{{Piskunov} \& {Kochukhov}(2002)}]{Piskunov02}
{Piskunov}, N. \& {Kochukhov}, O. 2002, \aap, 381, 736

\bibitem[{{Piskunov} \& {Rice}(1993)}]{Piskunov93}
{Piskunov}, N.~E. \& {Rice}, J.~B. 1993, \pasp, 105, 1415

\bibitem[{{Piskunov} {et~al.}(1990){Piskunov}, {Tuominen}, \&
  {Vilhu}}]{Piskunov1990}
{Piskunov}, N.~E., {Tuominen}, I., \& {Vilhu}, O. 1990, \aap, 230, 363

\bibitem[{{Semel}(1989)}]{Semel1989}
{Semel}, M. 1989, \aap, 225, 456

\bibitem[{{Solanki}(2003)}]{2003A&ARv..11..153S}
{Solanki}, S.~K. 2003, \aapr, 11, 153

\bibitem[{{Solanki} \& {Unruh}(1998)}]{1998A&A...329..747S}
{Solanki}, S.~K. \& {Unruh}, Y.~C. 1998, \aap, 329, 747

\bibitem[{{Vogt} \& {Penrod}(1983)}]{Vogt1983}
{Vogt}, S.~S. \& {Penrod}, G.~D. 1983, \pasp, 95, 565

\bibitem[{{Vogt} {et~al.}(1987){Vogt}, {Penrod}, \& {Hatzes}}]{Vogt1987}
{Vogt}, S.~S., {Penrod}, G.~D., \& {Hatzes}, A.~P. 1987, \apj, 321, 496

\end{thebibliography}

\end{document}